\documentstyle[aaspp4]{article}
\input epsf
\input hyperbasics

\def\spose#1{\hbox to 0pt{#1\hss}}
\def\simlt{\mathrel{\spose{\lower 3pt\hbox{$\mathchar"218$}}
        \raise 2.0pt\hbox{$\mathchar"13C$}}}
\def\simgt{\mathrel{\spose{\lower 3pt\hbox{$\mathchar"218$}}
     \raise 2.0pt\hbox{$\mathchar"13E$}}}
\def\mycaption#1{\vskip 0.1truecm
\rightskip=3truepc\leftskip=3truepc\baselineskip=11pt
        \noindent{\footnotesize#1}}

\begin{document}

\newcommand{\Phigb}{\Phi_{\gamma b}}
\newcommand{\RF}{{\cal{T}}}
\newcommand{\Aa}{{\cal A}_a}
\newcommand{\Ab}{{\cal A}_b}
\newcommand{\bg}{{b\gamma}}
\newcommand{\eal}{\!\!\! & = & \!\!\!}

\rightline{IASSNS-AST-96/6}
 
\title{ACOUSTIC SIGNATURES IN THE COSMIC MICROWAVE BACKGROUND}
\author{\vphantom{X}\href{http://www.sns.ias.edu/Main/members.html}{Wayne Hu}}
\affil{Institute for Advanced Study\\
School of Natural Sciences\\
Princeton, NJ 08540}
\authoremail{whu@sns.ias.edu}
\and
\author{\vphantom{X}\href{http://www-astro-theory.fnal.gov/Personal/mwhite/welcome.html}{Martin White}}
\affil{Enrico Fermi Institute\\
University of Chicago\\
Chicago, IL 60637}
\authoremail{white@oddjob.uchicago.edu}

\begin{abstract}
\noindent
\rightskip=0pt
We study the uniqueness and robustness of acoustic signatures in the 
cosmic microwave background by allowing for the possibility that they 
are generated by some as yet unknown source of gravitational perturbations.
The acoustic {\it pattern} of peak locations and relative heights predicted
by the standard inflationary cold dark matter model is essentially unique 
and its confirmation would have deep implications for the causal structure
of the early universe.  A generic pattern for isocurvature initial 
conditions arises due to backreaction effects but is not robust to 
exotic source behavior inside the horizon.  If present, the acoustic 
pattern contains unambiguous information on the curvature of the universe 
even in the general case. By classifying the behavior of the unknown 
source, we determine the minimal observations necessary for robust 
constraints on the curvature.  The diffusion damping scale provides an 
entirely model independent cornerstone upon which to build such a 
measurement.  The peak spacing, if regular, supplies a precision test.
\end{abstract}

\keywords{cosmology:theory -- cosmic microwave background}

\clearpage
\tableofcontents
\listoffigures
\clearpage
\rightskip=0pt
\section{Introduction}
\label{sec-intro}

Cosmic microwave background (CMB) anisotropies provide a unique window
into the early universe through which we receive information
on both the model for structure formation
in the universe 
(e.g.~\cite{EfsBonWhi,Goretal,OstSte,CriTur,Albetal,Duretal,DodGatSte})
and the background cosmology
(e.g.~\cite{Bonetal,HuSugSil,Jungman,Sel,ScoWhi}).
In the simplest models for structure formation, based on the gravitational
instability of initial density perturbations, the acoustic signature in the
anisotropy power spectrum provides a clean and unambiguous means of measuring
all of the parameters of a Friedman-Robertson-Walker cosmology.
In particular, it offers
a standard ruler with which to make a classical test for curvature 
in the universe (\cite{DZS,SugGou,KamSS,HuWhi}).  
However, structure formation may have proceeded
by a more complicated route.
Recent investigations have begun to probe the acoustic signature in
texture (\cite{CriTur,Duretal}) and string (\cite{Albetal,Magetal})
models for structure formation.  Their predictions differ strongly from the
standard inflationary case and suggest that perhaps without prior knowledge
of the correct model for structure formation, the information contained in the
acoustic signature cannot be extracted.
In this paper, we focus on two general questions:  does the acoustic signature
uniquely specify the model for structure formation? how robust is a measurement
of the curvature to changes in the underlying model?

As discussed in the appendix, 
the question of uniqueness is especially interesting in the case of the
standard inflationary paradigm.  Inflation is the unique {\it causal}
mechanism for generating correlated curvature perturbations
above the horizon (\cite{Lid}).
All other causal mechanisms generate significant curvature perturbations only
near horizon crossing.  We will refer to these alternate possibilities as
{\it isocurvature} models.
A unique signature of superhorizon curvature fluctuations can be used as a
test of inflation.  By generalizing the external source formalism of
Hu \& Sugiyama (1995a,b; hereafter \cite{HSa,HSb}) to include backreaction
effects in \S \ref{sec-physical}, we find that a large class of isocurvature models carry
a distinct acoustic signature that can be easily distinguished from the
inflationary case, independent of the curvature and other cosmological
parameters.  
The distinction lies in the gross properties of the spectrum, not
small or subtle shifts in the peaks heights which require very high resolution
measurements to see.
As shown in \S \ref{sec-signatures}, tell-tale features include the harmonic series of
peak locations, the alternating relative peak heights due to baryon drag,
and the diffusion damping tail.
From this study, we conclude in \S \ref{sec-inflation} that {\it the standard inflationary
model with big bang nucleosynthesis (BBN) baryon-to-photon ratio bears an
essentially unique signature}.
 
If acoustic oscillations are present in the CMB, as is the case in all but
models with significant reionization (see e.g.~\cite{Pee,EfsBon,Couetal}) 
or late formation of perturbations (\cite{Jafetal}),
their signature will provide information on the curvature.
The damping tail contains the most robust information, but its location alone
can constrain but not precisely fix the curvature.  
Additional restrictions such as a standard recombination history and a near
BBN baryon content are required to make this a sensitive probe of the
curvature.
Furthermore, in models where the acoustic signature is sufficiently
regular, the spacing of the peaks provides a precision test of the 
curvature even if the baryon content is anomalously low or somewhat high.  
It can also be combined with the damping tail to   
discriminate against truly exotic models. 
In \S \ref{sec-curvature}, we discuss specifically 
what regularities must be observed before the curvature 
can be unambiguously measured if the correct model for structure
formation is not assumed to be known {\it a priori}.

The outline of this paper is as follows.
The next section contains the derivation of our principle
results.  
Their impact on the features in the angular power spectrum of
the CMB is described in \S \ref{sec-signatures}.  
We discuss our main points in
\S\S \ref{sec-inflation} -- \ref{sec-curvature}, 
and present our conclusions in \S \ref{sec-conclusions}.

\section{Physical Processes}
\label{sec-physical}

This section develops the formalism behind our principle 
results and illustrates them in a series of concrete examples.
The lessons learned here will be applied in the later 
sections.  A summary of important conclusions is given
in \S \ref{ss-onesummary}, 
so that the bulk of this section may be skimmed on a first
reading.

Acoustic oscillations in the CMB are inevitable if gravitational 
potential perturbations exist during the period when the Compton
mean free path of a photon scattering off a free electron 
is much less than the horizon scale.  In this case, the photons and
electrons are tightly coupled.
Since Coulomb interactions couple the electrons to the baryons,
we refer to the system as a photon-baryon fluid.   Photon pressure
in the fluid resists gravitational compression 
and sets up acoustic waves in the system.  

Because the properties of the oscillator are determined by the background
while those of the gravitational forces are described
by the model for the perturbations, the acoustic signature provides
a unique opportunity to probe both the background cosmology and
the model for structure formation.  Here, we explore 
the evolution of acoustic phenomena under the influence of
an arbitrary source of gravitational perturbations.  This is employed
in \S \ref{sec-signatures} to determine the conditions under which signatures such as the peak
locations, relative heights, and damping tail may be considered robust.  
These signatures will be observable in the small scale CMB anisotropy 
if the tight coupling condition is satisfied during
the epoch immediately preceding the last scattering event.  
In particular, it holds for the standard thermal history where
recombination and hence last scattering occurs at redshift 
$z_*\sim10^3$.

\subsection{Fluid Equations}
\label{ss-fluid}

We start with the fundamental equations describing the dynamics of a
relativistic fluid. The physical interpretation and description of these
equations is given below.
The evolution of the photons and baryons in a metric perturbed
by density fluctuations in the $k$th normal mode 
is given 
in the Newtonian representation as (see e.g.~\cite{MukFelBra,MaBer})
\begin{equation}
\begin{array}{rcl}
\dot{\Theta}_0 \eal -{\displaystyle{k \over 3}} \Theta_1 - \dot\Phi, \\
\dot{\delta}_b \eal -kV_b - 3\dot\Phi,
\end{array}
\label{eqn:Continuity}
\end{equation}
for the continuity equations and
\begin{equation}
\begin{array}{rcl}
\dot{\Theta}_1 \eal k[\Theta_0 + \Psi - {1 \over 6}\Pi_\gamma] -
	\dot \tau(\Theta_1 - V_b), \\
\dot{V}_b \eal - {\displaystyle{\dot a \over a}}V_b + k\Psi + 
	\dot \tau(\Theta_1 - V_b)/R,
\end{array}
\label{eqn:Momentum}
\end{equation}
for the momentum conservation or Euler equations of the photons and baryons
respectively.  Here overdots are derivatives with respect to
conformal time $\eta=\int dt/a$, $R=3\rho_b/4\rho_\gamma$ is
the baryon-photon momentum density ratio, and $\dot{\tau}=x_e n_e \sigma_T a$
is the differential Compton optical depth with $x_e$ as the ionization
fraction, $n_e$ as the electron number density, and $\sigma_T$ as the 
Thomson cross section.  
The fluctuations are defined as $\Theta_0 =\Delta T/T=\delta_\gamma/4$
the isotropic temperature perturbation, $\Theta_1$ the dipole moment
or photon bulk velocity, $\Pi_\gamma$ the photon anisotropic stress
perturbation, $\delta_b$ the baryon energy density perturbation and 
$V_b$ the baryon velocity.  The gravitational sources are $\Phi$,
the perturbation to the spatial curvature, and $\Psi$, the Newtonian
potential.  In this gauge, these gravitational perturbations distort the 
metric as $g_{00} = -a^2 (1+ 2\Psi Q)$ and $g_{ij} = a^2(1+2\Phi Q)
\gamma_{ij}$, where $\gamma_{ij}$ is the three metric on a surface
of constant curvature and $Q$ is a plane wave $\exp(i {\bf k}\cdot{\bf x})$
in a flat geometry or more generally the $k$-eigenfunction of
the Laplacian (\cite{Wil}).  
The Einstein-Poisson equations thus relate them
to the matter fluctuations as (\cite{Bar})
\begin{equation}
\begin{array}{rcl}
(k^2 - 3K) \Phi \eal 4\pi G a^2 \sum \left[ \rho_i \delta_i + 3
\displaystyle{\dot a \over a}
(\rho_i + p_i)V_i/k \right], \\
k^2 (\Psi + \Phi) \eal -8\pi G a^2 \sum p_i \Pi_i,
\end{array}
\label{eqn:Poisson}
\end{equation}
where the sum is over particle species, 
the curvature $K=-H_0^2(1-\Omega_0-\Omega_\Lambda)$ and 
the Hubble constant $H_0 = 100 h $ km s$^{-1}$ Mpc$^{-1}$.
We will assume from now on that the relevant scales are far under the
curvature scale ($K/k^2\rightarrow0$). 
None of our main results are affected
by this assumption for reasonable values of $K$.
Note also that if the anisotropic stress $p_T \Pi_T = \sum p_i\Pi_i$ 
is negligible, $\Psi=-\Phi$.

Now let us examine the physical content of Eqs.~(\ref{eqn:Continuity})
and (\ref{eqn:Momentum}).
Photon number conservation relates changes in the temperature fluctuations
to the velocity divergence by a factor of ${1 \over 3}$ (since number density
is related to temperature by $n_\gamma \sim T^3$).
The $\dot{\Phi}$ term represents the dilation effect on the wavelength
of the photons.  Since the curvature perturbation ``stretches''
the spatial metric, changes in $\Phi$ gives rise to a dilation
effect of the same origin as the cosmological redshift.  
Notice the sign of this effect implies that the photons
will always {\it oppose} a change in the curvature, a point which
will be very important later.  Similar effects govern the baryon
continuity equation.  Since the fractional energy and number density 
fluctuations are equal for a non-relativistic particle, their
rate of change is given by the velocity divergence. The additional effect
due to the stretching of the volume from $\dot \Phi$ also 
implies a number density dilution of $3\dot \Phi$.

The expansion makes particle momenta decay as $a^{-1}$.  For the
photons, this is accounted for by the temperature
redshift; for the baryons, by expansion drag on the bulk velocity 
(the $\dot a /a$ term).
Gradients in the potential, $k\Psi$, generate velocity perturbations
by gravitational infall.  For the photons, infall is countered by
stress in the fluid, both isotropic (pressure) and anisotropic
(quadrupole moment).  The baryons however are effectively pressureless.
The photon and baryon equations are coupled by Compton scattering
(the $\dot{\tau}$ terms) which exchanges momentum between the fluids.
Since the momentum density of the fluid is proportional to $\rho+p$, 
conservation relates the scattering terms by the factor $R
=(p_b + \rho_b)/(p_\gamma +\rho_\gamma) \approx 3\rho_b/4\rho_\gamma$.
Scattering seeks
to equalize the bulk velocities $\Theta_1=V_b$ causing adiabatic evolution
of the density perturbations
 $\dot\delta_b = 3\dot\Theta_0$.

If the scattering is rapid compared with the travel time across a wavelength,
the momentum conservation Eqs.~(\ref{eqn:Momentum}) may be expanded in
powers of the Compton mean free path over the wavelength $k/\dot{\tau}$.  
By eliminating the baryon velocity, we obtain
the tight coupling approximation for the evolution of the photons
(\cite{PeeYu,HSa})
\begin{equation}
\begin{array}{rcl}
\dot \Theta_0 \eal -{\displaystyle{k \over 3}}\Theta_1 - \dot\Phi, \\
\dot \Theta_1 \eal - {\displaystyle{R \over 1+R}}
	  {\displaystyle{\dot{a}\over a}}\Theta_1
        + {\displaystyle{1\over 1+ R}} k\Theta_0 + k\Psi.
\end{array}
\label{eqn:TightCoupling}
\end{equation}
The quadrupole term $\Pi_\gamma = {\cal O}(k/\dot\tau) \Theta_1$ causes 
viscous damping and is treated
in \S \ref{ss-diffusion}.  It is a higher order correction
because scattering tends to isotropize the
photons in the baryon rest frame and suppresses the quadrupole.
From examining Eq.~(\ref{eqn:TightCoupling}), one can see that 
baryons decrease the efficacy of the pressure
and add an expansion drag term to the momentum equation.  The gravitational
infall term remains unaltered since 
its baryon analogue is identical.

\subsection{Gravitational Redshift Effects}
\label{ss-gravred}

Before proceeding with the main task of exploring the acoustic signatures,
let us review how gravitational effects manifest themselves in the CMB
(\cite{SacWol}).
As discussed in \S \ref{ss-fluid}, these are the ordinary redshift of a photon climbing
out in and out of potential wells and the dilation effect from changes in the
spatial metric.  If the metric fluctuations are generated by the density
fluctuations in the photon-baryon system itself, the Poisson 
equations~(\ref{eqn:Poisson}) tell us
that they are suppressed by a factor of $(k\eta)^{-2}$ with
respect to the temperature fluctuations inside the horizon.  Thus the
{\it self}-gravity  of the photon-baryon fluctuations generally
only is important outside the horizon $k\eta\ll1$.
However this is not necessarily true for metric fluctuations generated by an
external source. 
Gravitational redshift effects {\it can} significantly
alter the acoustic signature, and we must include them in the analysis even
on small scales.

Because temperature perturbations are observed only
{\it after} the photons 
have lost energy climing out of potential wells $\Psi$,
the ``effective'' temperature perturbation is given by $\Theta_0+\Psi$.
It is this quantity that we measure as a temperature fluctuation on
the sky.
It will be important in the following sections to consider the effective
temperature fluctuation rather than the intrinsic fluctuation $\Theta_0$.
The blueshift from infall into a constant gravitational well is exactly
cancelled by the redshift from climbing out.  Thus $\Theta_0$ may 
have a large but unobservable offset which is removed in
the effective temperature $\Theta_0 + \Psi$.  

On the other hand, the intrinsic temperature
evolves as $\dot \Theta_0 = - \dot\Phi$ above the horizon to
{\it oppose} changes in the spatial curvature.
Thus the effective temperature obeys the relation 
$\dot \Theta_0 + \dot \Psi = \dot \Psi - \dot \Phi \approx -2\dot\Phi$
above the horizon.  This yields a general description of
the gravitational redshift or Sachs-Wolfe effect (\cite{SacWol}),
\begin{equation}
[\Theta_0 + \Psi](\eta,k) = [\Theta_0 + \Psi](\eta_i,k) +  
  \left. [\Psi - \Phi](\eta,k) \right|^\eta_{\eta_i},
\label{eqn:SW}
\end{equation}
where $\eta_i$ is some initial epoch at which the fluctuations
were formed.  In particular, if the initial conditions are isocurvature
$\Phi(\eta_i,k) = 0$, and the initial temperature fluctuation is also
small, then the effective temperature becomes $\Psi-\Phi \approx -2\Phi$.
The photons are thus underdense in potential wells.
For adiabatic fluctuations, photons are overdense in potential wells
so that the effective temperature fluctuation is reduced.
As we shall see below, in the radiation-dominated era
$\Theta_0 + \Psi = {1 \over 2}\Psi$.  The change in the equation
of state through the matter-radiation transition causes a small
decay in the potential [see Appendix, Eq.~(\ref{eqn:ZetaEvol})] 
and brings the effective temperature in the matter
dominated limit to
$\Theta_0+\Psi={1\over3}\Psi$
(\cite{SacWol,HSa}). 

\subsection{Oscillator Equation}
\label{ss-oscillator}

In \cite{HSa}, a formalism was developed to calculate the response of the
photon-baryon fluid to metric fluctuations $\Psi$ and $\Phi$ which are
considered {\it external} to the fluid.  
Combining the two equations in (\ref{eqn:TightCoupling}), we obtain 
\begin{equation}
{d \over d\eta}(1+R)\dot\Theta_0 + {k^2 \over 3}\Theta_0 
  = -{k^2 \over 3}(1+R)\Psi - {d \over d\eta}(1+R)\dot\Phi.
\label{eqn:Oscillator}
\end{equation}
Conceptually, this equation reads: the change in momentum of the
photon baryon fluid is determined by a competition between the 
pressure restoring and the gravitational driving forces.
Below the sound horizon,
\begin{equation}
r_s(\eta) = \int_0^\eta  d\eta'\, c_s 
	= \int_0^{\eta} d\eta'\, {1 \over \sqrt{3(1+R)}}, 
\label{eqn:Sound}
\end{equation}
photon pressure resists gravitational compression and sets up
acoustic oscillations. 
Though conceptually useful, this approach has the
practical disadvantage that the photon-baryon contribution to
the metric fluctuations must be already known.
In general, it is unknown and we must 
break the metric
fluctuations up into pieces generated by the photon-baryon fluid
($\gamma b$) and by the external source ($s$), e.g.~from
dark matter, entropy fluctuations, defects, etc.
The Poisson equations (\ref{eqn:Poisson}) 
and the tight coupling condition $V_b=\Theta_1$ tell us
\begin{equation}
\Phi_\bg = -\Psi_\bg
	= 6 \Omega_\gamma H_0^2 (a k)^{-2} (1 + R)
	  [\Theta_0 + {\dot a \over a}{1\over k}\Theta_1] ,
\label{eqn:Poissonbg}
\end{equation}
for adiabatic fluctuations in the photon-baryon system where
$\delta_b = 3\Theta_0$.
As we shall see in \S \ref{ss-driven}, 
entropy fluctuations $S=\delta_b-3\Theta_0$ are simply
described as an external source since $\dot{S}=0$ in the tight coupling limit.
Thus Eq.~(\ref{eqn:Poissonbg}) represents no loss of 
essential generality.

As Eq.~(\ref{eqn:Poissonbg}) implies, the response of the photon-baryon
fluid is most easily solved in the so-called 
rest frame of the fluid (see 
Appendix).  
Here the temperature perturbation is
\begin{equation}
\RF = \Theta_0 + {\dot a \over a}{1 \over k} \Theta_1.
\label{eqn:RestFrame}
\end{equation} 
Combining Eqs.~(\ref{eqn:TightCoupling}) 
and (\ref{eqn:RestFrame}), it evolves under the equations
\begin{equation}
\begin{array}{rcl}
\left[ 1 + {\displaystyle{6 \over y^2}}(1+R) \right] \left[ \RF' 
- {\displaystyle{y'\over y}} {\displaystyle{1\over 1+R}} \RF \right] +
  {\displaystyle{1\over 3}}\left[ 1 - 3 {\displaystyle{y''\over y}}
+ 6\left({\displaystyle{y'\over y}} \right)^2 \right] \Theta_1
\eal {\displaystyle{y' \over y}}\Psi_s - \Phi'_s, \\
\Theta_1' +{\displaystyle{y'\over y}} \Theta_1  - 
  \left[ 1 - {\displaystyle{6 \over y^2}}(1+R)^2 \right]
  {\displaystyle{1\over 1+R}}\RF
\eal \Psi_s,
\end{array}
\label{eqn:Evolution}
\end{equation}
where primes are derivatives with respect to $x = k\eta$ with
$k$ fixed, and $y = (\Omega_\gamma H_0^2)^{-1/2}ak$.  
Notice that if $|\Theta_1| \simlt |\Theta_0|$, as is the case
in the oscillatory regime $xc_s\simgt 1$, Eq.~(\ref{eqn:RestFrame})
implies that $\RF = \Theta_0 [1 + {\cal O}(x^{-1})]$ and the two
representations become equivalent.
In the next few
sections, we will examine the implications of Eq.~(\ref{eqn:Evolution})
through a number of examples and approximations.

\begin{figure}[t]
\begin{center}
\leavevmode
\epsfxsize=3.5in \epsfbox{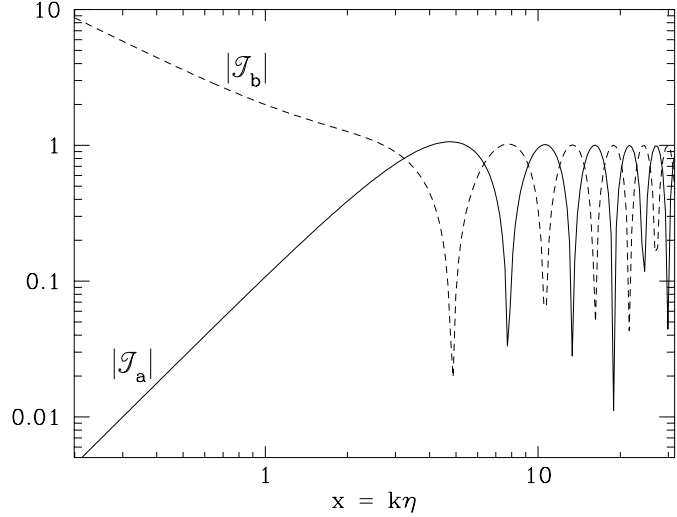}
\end{center}
\caption{Pure modes.}
\mycaption{In the absence of sources, the growing mode
of the rest frame temperature perturbation matches onto a cosine
acoustic oscillation inside the horizon, whereas the decaying mode
matches onto a sine oscillation.  The oscillator response to an
external source is constructed by Greens method from these homogeneous
solutions and transformed into the Newtonian frame with Eq.~(9).
}
\label{fig:hom}
\end{figure}

\subsection{Photon Backreaction at Early Times}
\label{ss-backreaction}

The evolution equation~(\ref{eqn:Evolution}) simplifies substantially
in the radiation-dominated epoch.
Early on, the baryon contribution is negligible and the fluid evolution is
dominated by the response of the photon perturbations to the source.  
The resulting photon-baryon density fluctuation feeds back into the
metric fluctuation.
Even though the universe may be just becoming matter dominated at 
last scattering, its early radiation-dominated legacy plays 
the dominant role.  
As we shall see, the processes that fix the amplitude of
the acoustic oscillation take effect mainly around horizon crossing
when the universe was still radiation dominated for the relevant
fluctuations.  

It is instructive to consider first the case in which the expansion
is photon-dominated.  
This neglects the neutrino and source contribution to the {\it background}
energy density, but does not fundamentally alter the results for the early
superhorizon evolution (see \cite{HSb}).
In this limit, $R\rightarrow0$, $y\rightarrow x$, and the evolution equations
(\ref{eqn:Evolution}) become
\begin{equation}
\begin{array}{rcl}
\RF' - \RF/x + \Theta_1/3  \eal
{\displaystyle{x^2 \over x^2 + 6}} \left[\Psi_s/x - \Phi_s'\right], \\
\Theta_1'+ \Theta_1/x - \left[1-6/x^2\right] \RF \eal \Psi_s,
\end{array}
\label{eqn:EvolutionSimple}
\end{equation}
or combining the two
\begin{equation}
\RF'' + {1 \over 3} \left[ 1 - {6 \over x^2} \right] \RF = {\cal S},
\label{eqn:SecondOrder}
\end{equation}
where the gravitational source is given by
\begin{equation}
{\cal S} = - \left[ {1\over 3} - {{12 \over (x^2 + 6)^2}} \right]\Psi_s
	- {{x(x^2+18)\over (x^2+6)^2}}\Phi_s' + {{x\over x^2 + 6}}\Psi_s'
	- {{x^2 \over x^2 + 6}}\Phi_s''.
\label{eqn:Source}
\end{equation}
If ${\cal S}=0$, this is simply the Bessel equation.
The homogeneous solutions are therefore (\cite{KodSas})
\begin{equation}
\begin{array}{rcl}
\RF_a(x) \eal 
- \cos (x / \sqrt{3})
+ (\sqrt{3} / x) \sin (x / \sqrt{3}) ,
\\
\RF_b(x) \eal 
               - \sin (x /\sqrt{3})
-(\sqrt{3} / x)\cos (x / \sqrt{3}) ,
\end{array}
\label{eqn:PureModes}
\end{equation}
with Wronskian $1/\sqrt{3}$.
These functions are plotted in Fig.~\ref{fig:hom} and may be considered as
the fundamental or {\it pure} modes of the photon-baryon fluid.  
Their limiting behavior as $x \rightarrow 0$ is
$\RF_a = x^2/9$ and 
$\RF_b = -\sqrt{3}/x$.
From Eq.~(\ref{eqn:RestFrame}), 
 the corresponding
limits for the Newtonian fluctuations are $\Theta_0 = 1/3$ and
$6\sqrt{3}/x^3$ and $\Phigb = 2/3$ and $-6\sqrt{3}/x^3$ 
for the two modes respectively.  As $x \rightarrow \infty$,
they become cosine and sine waves respectively for both the 
rest frame and Newtonian temperatures.

Although Eq.~(\ref{eqn:EvolutionSimple}) may easily be solved numerically,
the Green's method is more illuminating.   
Constructing the solution out of the pure modes, we find
\begin{equation}
\RF(x) = \Aa(x) \RF_a(x) + \Ab(x) \RF_b(x) ,
\end{equation}
with
\begin{equation}
\begin{array}{rcl}
\vphantom{\Big[}
\Aa(x) \eal \Aa(x_i) - \sqrt{3} \int_{x_i}^x \RF_b(x') {\cal S}(x') dx', \\
\Ab(x) \eal \Ab(x_i) - \sqrt{3} \int_{x_i}^x 
          \RF_a(x') {\cal S}(x') dx',
\end{array}
\label{eqn:Green}
\end{equation}
where $x_i$ is the initial epoch at which the perturbations
and hence
the constants $\Aa(x_i)$ and $\Ab(x_i)$ are fixed.
If the photon fluctuations are 
set to zero at $x_i$, then the answer can be read directly off the 
source behavior.  When 
$\RF_b \gg \RF_a$, source contributions stimulate the $\RF_a$ mode 
mainly.  An examination of Fig.~\ref{fig:hom}
would imply that all superhorizon effects from the source would
create a $\RF_a$ cosine mode.  
However, the initial fluctuations could be set up such 
that $\Aa(x_i)$ and $\Ab(x_i)$ exactly cancel the influence of the
source.  
This is exactly what occurs in the class of isocurvature models based
on balanced initial conditions.

To see this more clearly, let us express the
evolution 
in terms of $\Phi_{\gamma b}$, the curvature
perturbation generated by the photon-baryon fluid.
In the photon dominated limit, equation
Eq.~(\ref{eqn:Poissonbg}) gives $\Phi_{\gamma b} = 6\RF/x^2$ and
Eq.~(\ref{eqn:SecondOrder}) 
becomes
\begin{equation}
x^2 \Phigb'' + 4 x \Phigb' = - x^2 \Phi_s'' - 4 x \Phi_s' ,
\label{eqn:Compensationa}
\end{equation}
for $x \ll 1$.
The solution is
\begin{equation}
\begin{array}{rcl}
\Phi(x) \eal \Phigb(x) + \Phi_s(x) \\
\eal \Phigb(x_i) + \Phi_s(x_i)
   + {1 \over 3}[\Phigb + \Phi_s + (2 / x_i)\Theta_1](x_i)
	\left[ ( x_i / x)^3 - 1 \right],
\end{array}
\label{eqn:Compensation}
\end{equation}
where we have approximated $\Psi_s(x_i)=-\Phi_s(x_i)$ in 
rewriting the initial conditions in terms of $\Theta_1$.
Aside from a decaying mode, the photons evolve to keep the curvature
perturbation constant (c.f.~Eq.~(\ref{eqn:DecayingMode}) 
and \cite{VeeSte}).  
Thus, independent of the source behavior, if
$\Phi_\bg(x_i) = -\Phi_s(x_i)$ and $\Theta_1(x_i)=0$, compensation forces
$\Phi(x\ll1) \approx 0$.

While this argument only shows compensation for a photon dominated system,
the argument applies equally well for whatever the dominant dynamical
component is since $\dot{\delta}\propto -\dot{\Phi}$ for $x \rightarrow 0$.  
The argument does {\it not} strictly apply if the identity of 
this component,
i.e.~the equation of state, changes.  For example, pressure fluctuations and
hypersurface warping can change the curvature through the matter-radiation
transition even if $x\ll1$. We discuss these points further in the Appendix,
and show that densities and hence temperature fluctuations are anticorrelated
with the total curvature outside the horizon for isocurvature initial
conditions (see also Eq.~\ref{eqn:SW}).  

\begin{figure}[t]
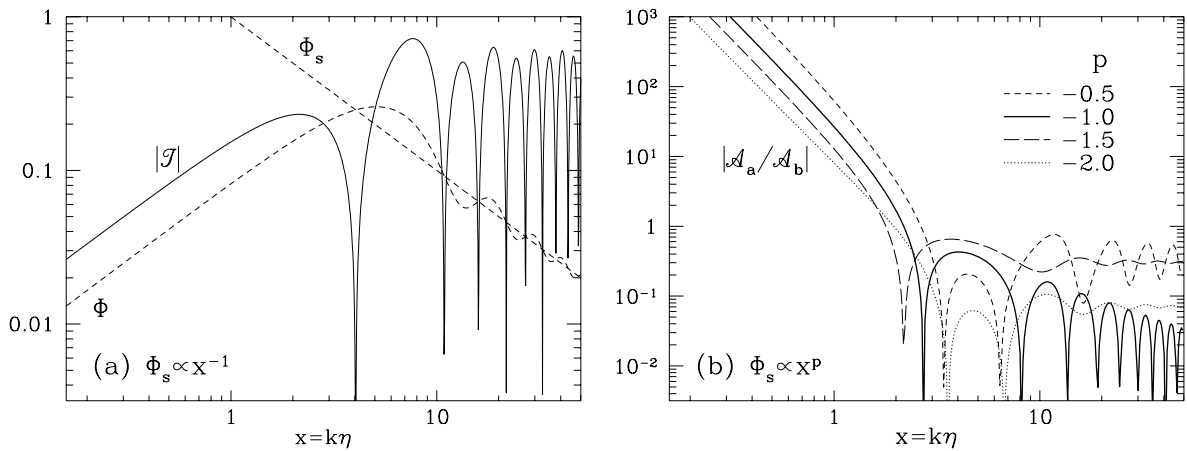

\begin{center}
\leavevmode
\epsfxsize=3.0in \epsfbox{f2a.epsf} \hskip 0.4truecm
\epsfxsize=3.0in \epsfbox{f2b.epsf}
\end{center}
\caption{Compensation and isocurvature fluctuations.}
\mycaption{(a) Baryon isocurvature model ($\Phi_s \propto x^{-1}$).  Outside
the horizon $x \simlt 1$, backreaction from the photons cancels the
contribution of the source to the curvature fluctuation.   Inside
the horizon, pressure prevents significant metric contributions
from the photon-baryon fluid and $\Phi \rightarrow \Phi_s$. (b)
For source functions $\Phi_s \propto x^{p}$, the ratio of $\RF_a$
to $\RF_b$ amplitudes decreases due to feedback.  This leaves the
acoustic oscillation mainly in the sine mode at $x \gg 1$.}
\label{fig:pib}
\end{figure}

\subsection{Driven Oscillations and Superhorizon Effects}
\label{ss-driven}

In this section, we will see how superhorizon compensation drives the
oscillator and stimulates one of the two pure modes of \S 
\ref{ss-backreaction}.  
Let us start with a simple and concrete example: the baryon isocurvature case.
Here, we begin at some initial epoch with an entropy fluctuation
$S(x_i)=\delta(n_b/n_\gamma) =\delta_b(x_i)-3\Theta_0(x_i)$.
Tight coupling implies that the number density fluctuation of the photons and
baryons, and hence $S$, remains constant [see Eq.~(\ref{eqn:Continuity})].
The entropy acts as an external source, which from 
Eq.~(\ref{eqn:Poisson}) contributes as $k^2 \Phi_s = 4\pi G a^2 
\rho_b S$, or rewriting this in the form of Eq.~(\ref{eqn:Poissonbg})
\begin{equation}
\Phi_s = -\Psi_s = {2 \over x^2} R S = {A \over x},
\label{eqn:PIBsource}
\end{equation}
where $A ={3\over2}(\Omega_bH_0^2)(\Omega_\gamma H_0^2)^{-1/2}k^{-1}S$.
Notice that the curvature perturbation implied by the source actually
{\it increases} as $x\rightarrow 0$.   For this model, the effective source
in Eq.~(\ref{eqn:Source}) reduces to the simple form
\begin{equation}
{\cal S} = -{1 \over 3}\Psi_s = {A \over 3} {1 \over x}.
\label{eqn:PIBS}
\end{equation}
Now let us assume the isocurvature condition at the initial epoch
$\Phi(x_i)=0$, or $\Phi_\bg(x_i)=-\Phi_s(x_i)$ and $\Theta_1(x_i)=0$ 
(see also Appendix).
By requiring continuity in $\RF$ and its first derivative, the initial
partition into pure modes in Eq.~(\ref{eqn:Green}) becomes
$\Aa(x_i)=-A/x_i$ and $\Ab(x_i) = (\sqrt{3}/54) A x_i^2$, yielding a
large contribution to the $\RF_a$ mode.   However, let us examine
the influence of the source term on the subsequent evolution,
\begin{equation}
\begin{array}{rcl}
\vphantom{\Big[}
\sqrt{3} \int_{x_i}^x  \RF_a(x'){\cal S}(x')dx'
 	\eal  -(\sqrt{3}/54) (x^2 - x_i^2) A , \\
\sqrt{3} \int_{x_i}^x  \RF_b(x'){\cal S}(x')dx' 
	\eal (x^{-1} - x_i^{-1}) A. \qquad (x \ll 1)
\end{array}
\label{eqn:PIBevol}
\end{equation}
Thus as the evolution progresses, the initially large $\RF_a$ 
contribution drops precipitously leaving mainly the $\RF_b$ or
sine mode at 
subhorizon scales $x \simgt 1$.  
Fundamentally, this is due to the feedback effect of
Eq.~(\ref{eqn:Compensationa}):
the photons oppose any change to the net curvature and evolve
to maintain the isocurvature condition.  In Fig.~\ref{fig:pib}, we show the
time evolution of $\RF$, the source curvature and the total 
curvature in the baryon isocurvature model.  By comparing
$\Phi$ to $\Phi_s$, notice that the 
feedback effect is only important outside
the horizon.  Another interesting quantity is the ratio of the
$\RF_a$ to $\RF_b$ amplitude shown in Fig.~\ref{fig:pib}b.  In this case,
the $\RF_a$ mode essentially disappears after horizon crossing 
leaving the acoustic perturbation in a pure sine mode ($p=-1.0$,
heavy line).  
In fact, the integral in Eq.~(\ref{eqn:Green}) takes on a simple 
asymptotic form (\cite{HSb})
\begin{equation}
\begin{array}{rcl}
\lim_{x \rightarrow \infty} \Aa(x) \eal 0, \\
\lim_{x \rightarrow \infty} \Ab(x) \eal A/\sqrt{3}.
\end{array}
\label{PIBaym}
\end{equation}
We also show in Fig.~\ref{fig:pib}b the behavior
when the source is generalized to   
$\Phi_s = -\Psi_s = A x^{p}$.  Again, the initially large 
$\RF_a$ mode is reduced as the superhorizon scale evolution progresses.

Even if the source has some superhorizon scale feature, 
i.e.~a maximum at $x\ll 1$, its
effect is mainly cancelled out.  
From Eq.~(\ref{eqn:Compensation}), notice that the photons always attempt
to counter the source.  The photon fluctuations would track
the rise and fall of a feature leaving an effect only from the boundary
conditions.  
Below the horizon however, 
compensation cannot occur due to the intervention
of photon pressure.  

\begin{figure}[t]
\begin{center}
\leavevmode
\epsfxsize=3.5in \epsfbox{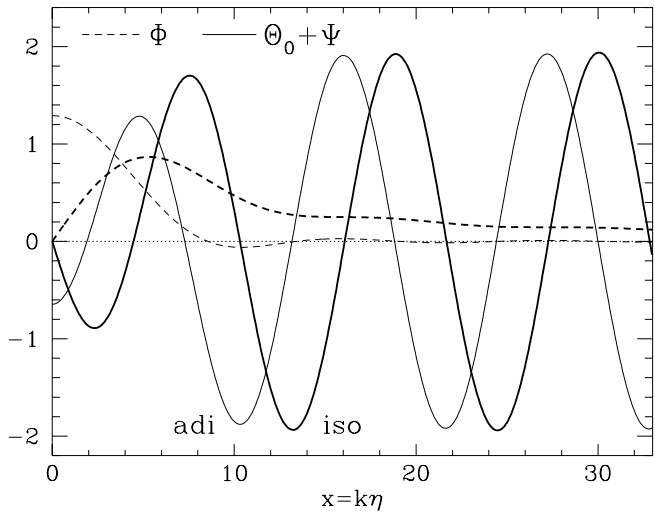}
\end{center}
\caption{Driven oscillations.}  
\mycaption{The self-gravity
of the photon-baryon fluid drives a cosine oscillation for adiabatic
initial conditions (thin lines)
and a sine oscillation (thick lines) for isocurvature initial
conditions.  The adiabatic model has $\Phi_s(x_i)=0$ and
$\Phi_{\gamma b}(x_i)\ne0$.  The isocurvature model is the baryonic
model of Eq.~(19).
The dashed lines show the full potential, the solid lines the effective
temperature.  In both cases the amplitude
in $\Theta_0+\Psi$ increase during the first few oscillations,
as described in \S \ref{ss-driven}.}
\label{fig:drive}
\end{figure}

How then does one obtain strong contributions to the $\RF_a$ cosine mode?
If superhorizon fluctuations are not fully compensated, 
Eq.~(\ref{eqn:Green}) implies
a large amplitude in $\RF_a$.  This {\it adiabatic} component
implies superhorizon curvature
perturbations initially: $\Phi(x_i) \ne 0$.  The simplest example is
the case of $\Phi_s = 0$ and $\Phigb(x_i) \ne 0$.  
In Fig.~\ref{fig:drive},
this case is contrasted with the baryon isocurvature case.  
Let us use these examples to gain further intuition about the feedback
mechanism.

From Fig.~\ref{fig:drive}, we see that in the isocurvature case,
compensation prevents a large curvature fluctuation from appearing outside of
the horizon regardless of the source.  However 
photon pressure, which becomes important around sound horizon
crossing, resists the accompanying rarefaction of
the photon fluid due to dilation.  
At this point, the fluid turns around and begins falling into 
the potential wells of 
the source enhancing the curvature fluctuation by its self-gravity.
Note the increase in the amplitude of the oscillation between the negative
maximum and the positive maximum.
As the photons resist further compression at the positive maximum, 
the self-gravity contribution to the potential fluctuation decays.  
This leaves the photon-baryon fluid in a highly compressed state and
increases the amplitude of the acoustic oscillation.  Thus
the self-gravity of the photon-baryon fluid essentially
{\it drives} the oscillator.  It provides a kick at each
of the first two turning points to enhance the oscillation.

A similar analysis applies to adiabatic fluctuations.  Here the
initial curvature fluctuation is kept constant outside the horizon
by photon backreaction.  From \S \ref{ss-fluid} recall that the intrinsic
temperature fluctuation and gravitational potential partially cancel in
the effective temperature.
{}From the $x\ll1$ limit of Eq.~(\ref{eqn:PureModes}),
$\Theta_0+\Psi_{\gamma b}={1\over 2}\Psi_{\gamma b}<0$.
At horizon crossing, the fluid begins to compress itself due to its 
self-gravity and the effective temperature reverses sign.  As pressure 
begins to stop the compression, the potential decays.  Again the
fluid is left in a highly compressed state, and self-gravity acts
as a driving term.
The dilation due to $\Phi$ (see discussion following
Eq.~(\ref{eqn:Poisson})) doubles the effect of infall from $\Psi$,
so that the end amplitude is
${1\over 2}\Psi_{\gamma b}(x_i)-2\Psi_{\gamma b}(x_i)\approx
-{3\over 2}\Psi_{\gamma b}$.  

A difference in phase between the two modes arises since
gravitational infall which initiates the chain of events leading
to the adiabatic acoustic mode only takes effect inside the horizon,
whereas the dilation effect that begins the isocurvature chain
of events is already occurring outside the horizon.  The end 
result is that the self-gravity of the photon-baryon fluid drives
a cosine oscillation for adiabatic initial conditions and a 
sine mode for isocurvature initial conditions.  
This is not the most general case however, 
since self-gravity and thus the initial conditions only dominate
the behavior around or before horizon crossing.  We consider the behavior
after horizon crossing in the next section.

\begin{figure}[t]
\begin{center}
\leavevmode
\epsfxsize=3.5in \epsfbox{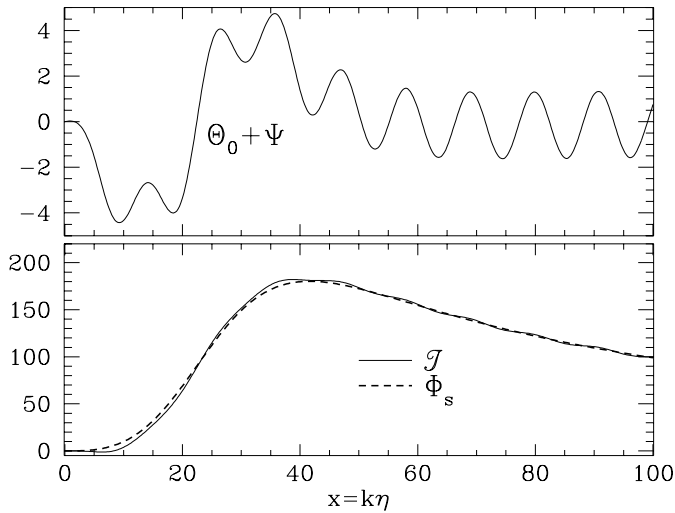}
\end{center}
\caption{Forced oscillations.}  
\mycaption{Inside the horizon,
the regulatory effects of photon-baryon self-gravity become ineffective.
A source that peaks at $x=k\eta\gg 1$ can produce complicated
forced oscillations in the fluid.  The intrinsic
temperature fluctuation $\RF\approx\Theta_0$ suffers a large
unobservable offset $-\Psi_s$ (bottom panel) which is removed by the
redshift in the effective temperature $\Theta_0+\Psi$ (top panel).
Notice that the fundamental period $kr_s = 2\pi$ or
$x\simeq 10.9$ is still clearly apparent for this slowly-varying source.}
\label{fig:force}
\end{figure}

\subsection{Forced Oscillations and Subhorizon Effects}
\label{ss-forced}

If the source is still effective inside the horizon $\Phi_s(x>1) \ge
\Phi_s(x=1)$, more complicated
behavior can result since the regulatory mechanism of photon feedback
is ineffective.  This is the case of so-called ``active perturbations''
(\cite{Magetal}).  The general Greens function solution can be
rewritten as
\begin{equation}
\RF(x) = [\Aa^2(x)+\Ab^2(x)]^{1/2}
	\sin \left\{ x/\sqrt{3} + \tan^{-1} \left[ \Aa(x)/\Ab(x) \right]
	\right\},
\label{eqn:DrivenSoln}
\end{equation}
for $x\gg 1$.  
The phase of the acoustic wave is in general time dependent
for subhorizon effects.  Furthermore, we cannot ignore the gravitational
redshift of the photons climbing out of the potential even inside the
horizon (see \S \ref{ss-gravred}).  
The effective temperature is
$\Theta_0+\Psi \approx \RF + \Psi$ if $x\gg 1$.
In Fig.~\ref{fig:force}, we plot an example based on a broken power law source
of the form
\begin{equation}
\Phi_s = {1 \over (\alpha x)^{-p_1} + x^{p_2}}.
\label{eqn:SourceForm}
\end{equation}
Here $p_1 = 3$, $p_2 = 1$ and $\alpha = 0.01$ such that the source peaks well
below the horizon scale.  Initial conditions do not play a major role here
since the fluctuation is generated inside the horizon.  
By considering only the effective temperature $\Theta_0 + \Psi$, we remove the
large unobservable offset from gravitational blueshift in $\RF$ (see \S \ref{ss-gravred}).
Note that it is important to include the Sachs-Wolfe contribution to obtain
a reliable measure of the temperature anisotropy at last scattering.
This is especially true when the anisotropies are ``sourced'' such that
metric fluctuations remain substantial inside the horizon, as in some
defect models.  Neglecting this effect, for example by use of the 
oft-employed synchronous-gauge photon density,
can lead to erroneous conclusions
about the shape of the CMB anisotropy power spectrum. 
Should the potentials have significant time variability after 
last scattering it is also necessary
to include the ISW effect, as this can lead to significant distortions of
the peaks, notably the first peak.

Even this smooth, well-behaved source leads to complicated structure in the
acoustic oscillation.  Furthermore, after the source turns off and $\Aa$ and
$\Ab$ become constant in time, the oscillator is left with a phase shift
described by Eq.~(\ref{eqn:DrivenSoln}) that does not necessarily correspond
to one of the pure modes.
It is important to note that even in this case, the temperature perturbation
is {\it anti}correlated with the source fluctuation immediately after horizon
crossing.  This is due to compensation from feedback around horizon crossing
and will be important for the question of robust distinctions between models.
We have found that this anticorrelation persists even if the source changes
sign outside the horizon or near horizon crossing.

It is of course impossible to quantify all the possibilities that arise from
the arbitrary action of a source well inside the horizon.
However Fig.~\ref{fig:force} illustrates two general points.
Large metric and correspondingly even larger density fluctuations are required
for subhorizon forcing to dominate over driving effects at horizon crossing.
Compare the vertical scales of the upper and lower panels of
Fig.~\ref{fig:force}.  There are two effects that make forcing around horizon
crossing easier than well inside the horizon.  The first is that for $x\gg1$ a
large metric perturbation requires a very large density perturbation, since
the Poisson equation (\ref{eqn:Poisson}) gives
$\Phi_s = (\rho_s/\rho_T) {\cal O}(\delta_s/x^2)$.  
Also, from the Euler equation, pressure perturbations are more important
than metric perturbations at large $x$, once the artificial offset due to
the $k\Psi$ term has been removed by considering the effective temperature.
From Eq.~(\ref{eqn:Oscillator}), the effective temperature evolves
according to
\begin{equation}
(\Theta_0 +\Psi)'' +
{1\over 3}       \left( \Theta_0+\Psi \right) = {\Psi''}-{\Phi''}
\label{eqn:EffTempEvol}
\end{equation}
The source is thus $\approx -2\Phi''$, and if constant, produces a temperature
shift of $6\Phi''$.
Exotic spectra of this type require extreme conditions.
More plausible is a model whose forced effects dominate only around $x\approx1$
where feedback effects still play some role.
As we shall see in \S \ref{ss-locationsf}, this has consequences for the coherence of the
resulting oscillations.

The second point is that in spite of the gravitational forcing the natural
period of the oscillation, here $x=2\sqrt{3}\pi$ is still apparent in the
harmonic structure.  If the source $\Phi''$ is slowly varying in time, so is
the phase shift.
In Fig.~\ref{fig:force}, the spacing between the peaks is still regular
and corresponds to the natural period despite the slowly varying offset.
Thus this acoustic signature is only lost if the metric fluctuations 
are both extremely {\it large} and {\it rapidly varying} inside 
the horizon.

If such an exotic spectrum is measured in the CMB, we would have concrete
evidence that the mechanism responsible for structure formation requires new
physics.  Until the spectrum is measured however, we can only consider the
broader implications.  
The possibility of driven sub-horizon effects raises two questions that we
will address in \S\S \ref{sec-inflation} -- \ref{sec-curvature}.  
Are there any unique signatures 
of the pure cases that distinguish them from
these more exotic scenarios?  How does the existence of such exotic models
affect our ability to measure quantities such as the curvature of the universe?
To help answer these questions, we need to understand two additional acoustic
effects that can provide clues to unraveling the spectrum.

\begin{figure}[t]
\begin{center}
\leavevmode
\epsfxsize=3.5in \epsfbox{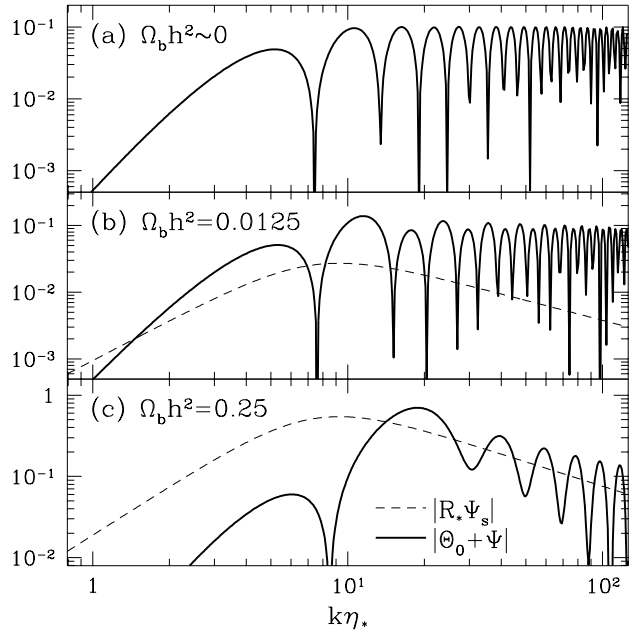}
\end{center}
\caption{Baryon drag and relative peak heights.}
\mycaption{Baryons displace the zero point of the oscillation by $- R\Psi$.
For a near BBN baryon content (center panel),
the displacement is smaller than the oscillation itself leading to
alternating peak heights in the rms.  For a much larger baryon content,
the photons oscillate around a strongly displaced zero point.}
\label{fig:drag}
\end{figure}

\subsection{Baryon Drag}
\label{ss-drag}

Up until this point, we have neglected the effect of the baryons
in the photon-baryon fluid.  This is appropriate for the early 
evolution and reveals the qualitative structure of the acoustic
signal even at recombination since
$R(\eta_*)=3\rho_b/4\rho_\gamma |_{\eta_*}\approx 31.5\Omega_b h^2\sim30\%$.
Nevertheless, there is an important acoustic signature associated with the
baryons if $\Omega_b h^2$ is near or greater than this BBN value (HSa,
and Hu \& Sugiyama 1996, hereafter HSc).  
To include its effects properly, we must solve the full
Eq.~(\ref{eqn:Evolution}).
However it is instructive to examine the qualitative origin of the
effect first.  
Eq.~(\ref{eqn:Oscillator}) tells
us that the baryons contribute to the inertial and gravitational
mass of the fluid.  Thus with a higher baryon content, gravity
can compress the
fluid more strongly inside the potential well $\Psi < 0$.  
In the limit that $x \gg 1$ and 
we observe for only a time short compared with changes in 
$R$ and the source, the solution to the oscillator equation~(\ref{eqn:Oscillator}) is
\begin{equation}
\Theta_0 + \Psi = C_1 \cos(kr_s) + C_2 \sin(k r_s) - R\Psi.
\label{eqn:DragForm}
\end{equation}
The last term represents the baryon drag on the photons and
displaces the zero point of the oscillation.  Since $-R\Psi > 0$
inside a potential well, it enhances the compression and suppresses
the rarefaction stage of the acoustic oscillation in the potential
well.  The crucial point is that the presence of baryons allows
us to distinguish between the two stages
from the observable rms deviation $|\Theta_0 + \Psi|$.  Fig.~\ref{fig:drag}
shows an example based on the source in Eq.~(\ref{eqn:SourceForm})
with $p_1 = 2$, $p_2 = 1$ and $\alpha = 0.05$.  Here we take the
$k$-spectrum at recombination for a baryon content of
$\Omega_b h^2=0,0.0125,0.25$ where $R_*=R(\eta_*)=0,0.38,7.6$.
The neutrino and CDM densities have been set to zero 
for simplicity.  Notice that if baryon drag is significant
but not dominant (middle panel) then it will
modulate the peaks into a pattern of alternating peak heights.
If it dominates (bottom panel), the rms fluctuations no 
longer possess zero crossings but oscillate around some large
d.c.~offset. Either pattern is quite distinctive and as we shall
see can separate the adiabatic from isocurvature scenarios in 
{\it all} cases save those with
$|R\Psi| \ll |\Theta_0+\Psi|$ at last scattering. 

\subsection{Photon Diffusion}
\label{ss-diffusion}

Finally, photon diffusion leaves a robust signature by providing a cut off
scale to the acoustic oscillations that is independent of the source
fluctuation.  As the diffusion length passes the wavelength, acoustic
oscillations are exponentially damped (see \cite{Sil,Wei}).  
Physically this occurs since diffusion erases temperature 
differences across a wavelength and causes 
viscosity (or anisotropic stress) in
the fluid.  
Anisotropic stress, or the quadrupole, is generated from the
free streaming of a dipole fluctuation $\Theta_1$.  
As photons from crests and troughs of the original velocity
perturbation meet, their Doppler shifts create a quadrupole
temperature pattern.  This transfer of power from the dipole to
the quadrupole is but a manifestation of a general tendency.
Streaming transfers power to higher angular moments since 
the original temperature
fluctuation subtends a smaller and smaller angle as seen by a distant
observer.
However in the tight coupling limit, streaming is collisionally
suppressed
by the factor $\dot\tau/k$ (the optical depth through
a wavelength) such that in the tight coupling limit, the anisotropic
stress is approximately
(see HSc, Eq.~A8)
\begin{equation}
\Pi_\gamma = {8 \over 5}(k/\dot\tau) f_2^{-1} \Theta_1,
\label{eqn:AnisoStress}
\end{equation}
where $f_2 = {3 \over 4}$ (\cite{Kai}).  
Other approximation commonly employed
are $f_2 = {9 \over 10}$ for unpolarized radiation (\cite{Chi})
and $f_2 = 1$ for further neglecting the angular dependence of
Compton scattering (\cite{Wei},HSa).  
From Eq.~(\ref{eqn:DragForm}),
it is natural to try a solution of the form $\exp i\int \omega\, d\eta$
for both $[\Theta_0+(1+R)\Psi]$ and $\Theta_1$.  
Heat conduction, proportional to $V_b-\Theta_1$, is described by iterating
the baryon Euler equation (\ref{eqn:Momentum}) to second order,
\begin{equation}
V_b - \Theta_1 = - \dot\tau^{-1} R [i\omega\Theta_1 - k\Psi] -
\dot\tau^{-2} R^2 \omega^2 \Theta_1.
\label{eqn:Heat}
\end{equation}
Here we have ignored changes on the order of an expansion time compared
with those at the oscillation period.
Combining Eqs. (\ref{eqn:AnisoStress}) and (\ref{eqn:Heat}) in the   
photon Euler equation (\ref{eqn:Momentum}), we obtain the 
dispersion relation
\begin{equation}
\omega = \pm {k c_s} + i {1 \over 6} k^2 \dot \tau^{-1}
\left[ {R^2 \over (1+R)^2} + {4 \over 5} f_2^{-1} {1 \over 1+R} 
	\right],
\label{eqn:Dispersion}
\end{equation}
where recall that the sound speed $c_s = 1/\sqrt{3(1+R)}$.
It follows that acoustic oscillations are damped as (HSc)
\begin{equation}
\Theta_0 + \Psi = [\hat \Theta_0 + \Psi] e^{-[k/k_D(\eta)]^2}
		 + R\Psi (e^{-[k/k_D(\eta)]^2} - 1),
\end{equation}
where $\hat \Theta_0$ is the acoustic signal described
by Eq.~(\ref{eqn:TightCoupling}) and the damping wavenumber is
\begin{equation}
k_D^{-2} = {1 \over 6} \int d\eta\ {1\over\dot{\tau}}\,
{R^2 + 4f_2^{-1}(1+R)/5 \over (1+R)^2}.
\label{eqn:DiffusionLength}
\end{equation}
To order of magnitude, the diffusion length is the geometric mean of the
Compton mean free path $\dot \tau^{-1}$ and the horizon $\eta$, as expected
of a random walk.
Aside from the residual baryon drag effect, photon diffusion leaves
a distinctive damping signal in the CMB that is only dependent on the
cosmological background.  Because the damping is exponential, no oscillations
survive at $k\gg k_D$ regardless of the source.  

One complication arises however.  As recombination progresses, the ionization
fraction and hence the differential optical depth $\dot{\tau}$ decreases.
The corresponding increase in the damping length can still be approximated in
the tight coupling limit if $\dot{\tau}/k_D(\eta_*)\sim k_D\eta_*\gg1$.
In this case, the diffusion length passes the wavelength while 
tight coupling holds and the damping can be calculated semi-analytically
from the known ionization history (see HSc).  If this condition is not
satisfied, as is the case for extremely low $\Omega_b h^2$ or
reionized models 
with a long
Compton mean free path at recombination, the damping must be calibrated
numerically (Hu \& White, in preparation).

\subsection{Summary}
\label{ss-onesummary}

In summary then, we have shown that the evolution of fluctuations acts to
resist changes in the spatial curvature, whenever the pressure can be
neglected (outside the horizon).
Schematically, this is because the radiation or matter in a growing
curvature
perturbation redshifts faster than average (c.f.~cosmological redshift,
as in \S \ref{ss-fluid}) reducing the local energy density and hence 
stabilizing the curvature
perturbation. It is also a consequence of causality as shown in 
the Appendix.

The behavior of the acoustic oscillations due to 
sources outside the horizon thus depends
only on the initial conditions: whether or not there are initial
uncompensated curvature perturbations.
The compensation by the radiation also implies that the self-gravity
of the photons produces an important feedback mechanism.  
The result is that sources arising before or near
horizon crossing stimulate either the cosine or the sine mode of the
oscillation, not an arbitrary admixture.

The fact that the compensation mechanism leads to rarefactions in the photon
density at horizon crossing for isocurvature fluctuations but compressions for
adiabatic fluctuations inside the potential well is an extremely robust
conclusion.
Of crucial importance is it leads to an observable effect in models with
$\Omega_b h^2$ of a few percent or greater due to the drag induced by the
baryons.  Photon diffusion also leaves a distinct damping tail that is
entirely independent of the model for the fluctuations but sensitive to
the thermal history.

\section{Acoustic Signatures}
\label{sec-signatures}

Having studied the acoustic oscillations in the last section, let us now
consider what signatures the effects summarized in 
\S\ref{ss-onesummary} leave in the
angular power spectrum of CMB anisotropies.
Acoustic oscillations are frozen in at last scattering $\eta_*$
when the Compton optical depth to the present drops below unity. 
The evolutionary properties of the acoustic phenomena 
before last scattering 
influence the peak locations, heights, and damping tail.  
These will be used in the
next sections to devise tests of the model for structure formation
and the background cosmology.

\subsection{Peak Locations for Driven Oscillations}
\label{ss-locationsd}

At last scattering, each $k$-mode is caught at a different phase 
in its oscillation.  
Of course, one can adjust the magnitudes in $k$ 
by choosing different relative weights for the $k$-modes
at the initial conditions.  We assume that any such weighting is
not in itself oscillatory in $k$.  
Under this single assumption, the acoustic pattern of 
peaks is robust to the details of the model even though their
absolute heights are not. 
These oscillations on the last scattering surface
are viewed today as anisotropies on
the sky.  
Thus the $k$-space power spectrum is projected onto an 
angular power spectrum.  The angular correlation function 
is broken up into $\ell$ Legendre moments which represent the 
power on the scale $\theta \sim \ell^{-1}$.  The power spectrum
is usually denoted as $\ell (\ell + 1)C_\ell$ where $C_\ell$ is
the ensemble average of the squared multipole coefficients.  

As 
we shall see in \S \ref{sec-curvature} 
[(see Eq.~(\ref{eqn:Projection})], 
since the projection of $k$ onto $\ell$ is highly
dependent on the curvature, knowledge of the physical scale of acoustic
features allows a sensitive probe of the curvature.  
The physical scale of the peaks is related to the sound horizon 
$r_s = \int c_s \, d\eta \approx \eta/\sqrt{3}$ as $R \rightarrow 0$
or more generally
\begin{equation}
r_s = {2\over\sqrt{3}}(\Omega_0 H_0^2)^{-1/2} 
	\sqrt{ a_{\rm eq} \over R_{\rm eq} } \ln
	{ \sqrt{1 + R} + \sqrt{R + R_{\rm eq}} \over 1 + \sqrt{R_{\rm eq}}},
\end{equation}
assuming the universe was radiation dominated before $a_{\rm eq}$. 
The exact relation of the peak scale to the sound horizon 
depends on the nature of the fluctuations and supplies a
test of the model (see \S \ref{ss-reconstruction}).   
Let us now examine this relationship more carefully.

Once the source becomes ineffective $|\Theta_0| \gg |\Phi_s|$,
the partition into pure mode amplitudes $\Aa$ and $\Ab$ of
\S \ref{ss-backreaction} becomes
time independent and the 
acoustic oscillations settle into a pure sinusoidal form,
$\sin\phi(\eta,k)$, with phase
\begin{equation}
\phi = k\,r_s + \tan^{-1} (\Aa / \Ab).
\end{equation}
For sources which peak before horizon
crossing, backreaction effects create
a two state system (see \S \ref{ss-driven}).   
If the fluctuations are not perfectly compensated
at the initial conditions, $\Aa \gg \Ab$.  Compensation tends to 
create a situation where $\Ab > \Aa$.  

Robust predictions arise for the location
of the peaks in the $k$ or $\ell$ space power spectra.  
Since the phase is dependent only on the nature of the initial 
fluctuation, not on the detailed behavior or the source in time
or $k$, all such models give definite predictions for the
peak locations. 
\begin{equation}
\phi(k_m) = (m-1/2)\pi,
\label{eqn:PeakPhase}
\end{equation}
for integer $m$.  
Furthermore, the harmonic series of peak locations is 
independent of the background quantities that fix the sound horizon and
the angle it subtends on the sky.  The generic adiabatic prediction
is that peak locations follow a $1 : 2 : 3 \cdots$ series in $k$ or $\ell$.
Isocurvature models tend toward a $1 : 3: 5 \cdots$ pattern (\cite{HSb}).  
Even if $\Aa$ and $\Ab$ are comparable, as would be the case if a small
(finely tuned) uncompensated fluctuation remained in the initial condition,
the pattern of peaks uniquely determines the phase shift.  
This might be the case if the initial conditions contain a balanced
amount of coherent adiabatic and isocurvature fluctuations.  
In most physical examples however (see e.g.~\cite{KawSugYan}), the processes
which generate the two types of fluctuations are statistically independent
and the two contributions are incoherent, i.e.~generated with different phases
in $k$.  In this case it is even more unlikely that the two contributions
would be balanced to give the same acoustic amplitudes.   
For incoherent superpositions, the phase is always determined by
the dominant component with a change in amplitude, not phase, caused by the
weaker component.  
Furthermore, the phase difference $\phi(k_m)-\phi(k_{m-1}) = \pi$
implies that regardless of the phase shift or amplitude
variation, the spacing of the peaks in
$k$ is given by
\begin{equation}
k_A = k_m  - k_{m-1} = \pi/r_s
\label{eqn:Kspacing}
\end{equation}
Thus the acoustic peaks possess both model information in the
ratios of their locations and model independent information for
the measurement of background parameters in the spacing between the
peaks.

\subsection{Peak Locations for Forced Oscillations}
\label{ss-locationsf}

Although definiteness in phase is a typical feature of acoustic
oscillations,
it is not necessarily
obeyed by models where the source is still active inside
the horizon.  In this case, the partition into pure modes through
$\Aa$ and $\Ab$
is time dependent.
There are two limiting cases worth considering.  
The source could be scale
free in its temporal behavior such that $\Phi_s(x,k)=G(x)F(k)$ 
where recall $x=k\eta$.
This occurs for example 
if the source behavior is correlated with horizon crossing
due perhaps to the onset of a causal mechanism.
As discussed in \S \ref{ss-forced}, 
this is in fact the most likely scenario
since metric perturbations that can overcome photon pressure
become increasingly hard to generate
inside the horizon.
In this case, the behavior as a function of $x=k\eta$
represents both the time evolution of a fixed $k$-mode and the
transfer function in $k$ for a fixed time, e.g.~at last scattering
$\eta_*$.  If $G(x)$ peaks around $x=1$ as is typical, we would expect
an irregular first peak followed by an increasingly regular but phase
shifted harmonic series.
If it peaks much after $x=1$, complicated acoustic behavior containing
detailed information about the source evolution would
result (see Fig.~\ref{fig:force}).
Even in this case, we can see from Eq.~(\ref{eqn:EffTempEvol}) that if the
source is slowly varying in time such that $G''''\ll c_s^2 G''$, 
the natural period
is still imprinted in the oscillations.  
In cases where the driving force is large, 
we expect only compressional phases
to be clearly visible as peaks in the rms. 
Rarefactions become rms troughs.  Using only the compressional extrema,
the peak spacing is then given by $2k_A = 2\pi/r_s$.

In the extreme case that the metric perturbations are generated
well after $x=1$, one might also expect stochastic
behavior in the source
(\cite{Albetal}).
Here the timing mechanism due
to the act of horizon crossing no longer serves to correlate the modes.
In this case, each $k$-mode receives a different set of impulses from the
source.  
The phase shift given by $\tan^{-1}[\Aa(x,k)/\Ab(x,k)]$ varies with
$k$. In the extreme limit of
rapid variation, the phase information is lost when summing over
$k$-modes to form the observable anisotropy.  
This washes out the oscillatory behavior.  
It is believed that a concrete example of this 
mechanism is given by the cosmic string scenario (\cite{Albetal})
although whether it is the dominant mechanism or not is currently
disputed.  We shall discuss what can be learned in the case where the
oscillations are washed out in \S \ref{sec-curvature}.  
Let us assume for the present that
oscillations will be observed.

\subsection{Peak Heights}
\label{ss-heights}

A wealth of information is stored in the peak heights.  Their
signature is more model dependent than the peak locations and provides
an excellent means of examining the fine details of the model.  
On the other hand, robust features that distinguish between general
classes of models are more difficult to isolate.  In this section,
we will examine {\it generic} features in the peak heights that may 
at least be used as clues for this purpose. 

Let us begin by examining the familiar scale-invariant adiabatic
case.  In the tight coupling limit, there are two effects
that determine the heights of the peaks: the driving force of
feedback and baryon drag. 
They are to a certain extent mutually exclusive.
As we have seen in \S \ref{ss-driven}, 
feedback boosts the oscillation amplitude above the initial conditions
$[\Theta_0+\Psi](\eta_i,k) = {1 \over 2}\Psi(\eta_i,k)$ by a factor of
$3$.  
Recall that the fluid first is compressed by its own self-gravity.
Photon pressure resists the compression, causing the photon-baryon
contribution to the potential to decay.  The fluid is thus 
released exactly in this highly compressed state into the acoustic
phase.
We show in the Appendix that the potential in the matter dominated limit 
$\Psi(\eta,k)={9\over10}\Psi(\eta_i,k)$, thus the boost represents a factor
of $5$ enhancement over the $\Theta_0 + \Psi = {1 \over 3}\Psi$ Sachs-Wolfe
tail. 
The inclusion of the neutrino background slightly lowers the amplitude
(see \cite{HSc}).
The driving effect does not occur if the potentials are dominated
by an external source such as cold dark matter.  Hence the prominence
of the acoustic oscillations increases if the universe is made more
photon-baryon or radiation dominated at last scattering.  

Baryon
drag enhances the compressional, here odd, peaks by a term of
${\cal O}(R\Psi)$.  This alters the peak heights to give
the distinctive alternating or offset oscillation pattern but 
becomes sub-dominant if the potential is intrinsically small.
Such is the case for driven oscillations 
since the potential $\Psi\approx \Psi_{\gamma b}$ and decays inside
the sound horizon. 

For isocurvature models, gravitational redshifts cause the effective
temperature above the horizon to be $\Theta_0 + \Psi \approx 2\Psi < 0$
(see \S \ref{ss-gravred}).  
It is also worthwhile to note that although the fluctuation outside
the horizon at last scattering is small in these models, this does
not imply that the observable anisotropy from those scales is 
correspondingly small.  Gravitational redshifts from the time dependent
potential continue to generate fluctuations between last scattering
and the present in the {\it same} manner.  This is generally called the
integrated Sachs-Wolfe (ISW) effect.  Thus an isocurvature model
which generates a scale invariant spectrum of curvature fluctuations
near horizon crossing yields a flat large scale anisotropy spectrum
just as in the familiar adiabatic case (\cite{BenSteBou,Couetal,PenSpeTur}).  

Superhorizon isocurvature evolution makes the photon-baryon fluid 
more and more rarefied inside
a potential well until photon pressure can successfully resist
rarefication.  Since the first feature is the turning point from
the superhorizon behavior, it is not prominent in comparison to 
the $2\Psi$ Sachs-Wolfe effect.  However, the fluid then begins infall into
the source wells.  The driving effect of the photon-baryon contribution
to the potential wells now proceeds as in the adiabatic
case to enhance the oscillation making the {\it second} feature
much more prominent than the first (see Fig.~\ref{fig:drive}).
Whereas the first peak has a height of order 
$2\Psi$ at $kr_s = \pi/2$ and is
small since compensation eliminates metric fluctuations
above the horizon, the second peak has a height of order $2\Psi$ at
$kr_s = 3\pi/2$, which is significantly larger since the photon-baryon
contribution adds to rather than cancels the source.
As in the adiabatic case, baryon drag contributes an ${\cal O}(R\Psi)$ term
that boosts the compressional phases.  However in the isocurvature case,
these are the {\it even} peaks since compensation demands that if the first
feature occurs near the horizon it represents rarefaction inside the
potential well.
Thus, the second isocurvature peak is lifted even higher with respect to the
first by baryon drag.

How robust are these general tendencies?  The prominence of the acoustic
oscillations compared with the large scale tail can be masked or altered
by the presence of tilt or features in the initial $k$-spectrum as well as
by other effects from tensor and vector metric perturbations
(e.g. \cite{CriTur}).
These effects are however unlikely to obscure the distinctive alternating peak
heights due to baryon drag.  Unfortunately, the baryon drag effect may be
difficult to observe if
$|R \Psi|(\eta_*)\ll |\Theta_0 + \Psi|(\eta_*)\sim
|\Psi|(\eta \approx 1/c_s k)$.
Even with high precision measurements, one must first remove the 
diffusion damping envelope in this case.  

\begin{figure}[t]
\begin{center}
\leavevmode
\epsfxsize=3.5in \epsfbox{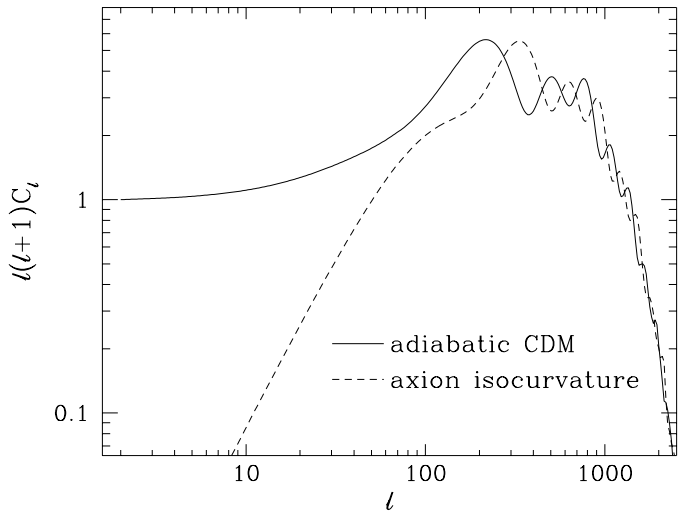}
\end{center}
\caption{Diffusion damping.}  
\mycaption{Although adiabatic
and isocurvature models predict acoustic oscillations in different
positions, they both suffer diffusion damping in the same way.
The damping length is fixed by background assumptions, here
$\Omega_0=1$, $h=0.5$, $\Omega_b=0.05$ and standard recombination.
These calculations were performed
using a full numerical integration of the Boltzmann equation with
the code of Sugiyama (1995) as were results in Figs.~7,8,10,11.
}
\label{fig:damp}
\end{figure}

\subsection{Damping Tail}
\label{ss-tail}

Unless baryon drag dominates over the acoustic oscillation, as in the case that
$|R \Psi|(\eta_*)\gg |\Theta_0+ \Psi|(\eta_*)$,
the damping leaves a clear signal in the CMB.
To demonstrate this robustness to changes in the model for the gravitational
source, we compare in Fig.~\ref{fig:damp} the anisotropy power spectra of a
standard inflationary model with scale invariant curvature fluctuations at
horizon crossing and a similar axionic isocurvature model.  The background
parameters are set to be equal at $\Omega_0 = 1.0$, $h=0.5$ and
$\Omega_b = 0.05$.  Notice that the damping behavior is independent
of the nature of the fluctuations.

The angular location of the damping tail is highly sensitive to the curvature
which dominates projection effects (see \S \ref{ss-dampingtail}) 
but also to  the thermal history which sets
the maximum diffusion scale, the baryon content $\Omega_b h^2$ which sets the
mean free path of the photons, the matter content $\Omega_0 h^2$ 
which sets the horizon scale.  
A measurement of the damping scale alone would fix the combination of
these quantities defined by Eq.~(\ref{eqn:DiffusionLength}).
More specifically, if last scattering occurs due to standard recombination,
the wavelength $k_D$ at which the acoustic amplitude falls to $e^{-1}$
of its original amplitude can be fit by semianalytic techniques
(HSc, Eq.~E6) to $\sim10\%$.
Unfortunately, a measurement of $k_D$ cannot break the degeneracy in these
quantities.  If $\Omega_b h^2$ is constrained by BBN and $\Omega_0 h^2$ 
by dynamical mass and Hubble constant measurements, 
it would provide interesting constraints
on the curvature of the universe as we shall show in 
\S\ref{ss-dampingtail}.  
Furthermore with a measurement of the acoustic peaks themselves, the degeneracy
can potentially be removed in many models for the source fluctuations
(see \S\ref{ss-reconstruction}).  

\section{Uniqueness of the Inflationary Spectrum}
\label{sec-inflation}

The inflationary paradigm is the front running candidate
for a mechanism of fluctuation generation in the early universe: the
perturbations in density which are the precursors of galaxies
and CMB anisotropies today.
Are there unique signatures of inflation that can validate the
paradigm?
For a long time it was thought that a nearly scale-invariant spectrum of
anisotropies would provide evidence for inflation (\cite{KolTur}).
But then it was realized that models of structure formation based on
topological defects also naturally formed scale-invariant spectra
(see \S \ref{ss-heights} and \cite{BenSteBou,Couetal,PenSpeTur}).
From the other direction, finding that the universe had non-vanishing
spatial curvature once seemed like a way to ``disprove'' inflation, however
recently it has been shown that inflation could survive such a revelation
(\cite{BGT,YamSasTan}).

Currently the most popular means of ``proving'' inflation is to test the
consistency relation between the tensor (gravity wave) and scalar 
(density) modes.   
By measuring the detailed shape of the anisotropy spectrum, one may 
infer the relative amplitudes of scalar and
tensor perturbations.  The ratio of tensor to scalar
contributions is proportional in slow-roll inflation to the slope of the
tensor spectrum (\cite{LidLyt,Davetal,TurWhi}).
However this method of proof requires a precise determination of the tensor
to scalar ratio, which in turn requires a sizable fraction of the 
anisotropy be contributed from tensor modes (\cite{WKS,KnoTur}).
What one really desires is a test based on the most basic ideas of the
inflationary scenario that will be both observationally
feasible and able to survive the
ingenuity of model builders.

The key feature of inflation for our purposes is that it provides a mechanism
of connecting parts of the universe at early times which are
currently space-like separated.  In fact, it can be shown that inflation is
the unique {\it causal} mechanism for correlating 
curvature perturbations on scales larger than the horizon
(\cite{Lid,HuTurWei}).  
The possibility of a white noise spectrum of
superhorizon curvature perturbations is discussed in the Appendix,
but is already observationally ruled out as a means of structure formation.
The question therefore arises, are there unique consequences of 
such super-horizon
curvature perturbations?  If so, their observation would 
provide a ``proof''
of the inflationary paradigm.

\begin{figure}[t]
\begin{center}
\leavevmode
\epsfxsize=3.5in \epsfbox{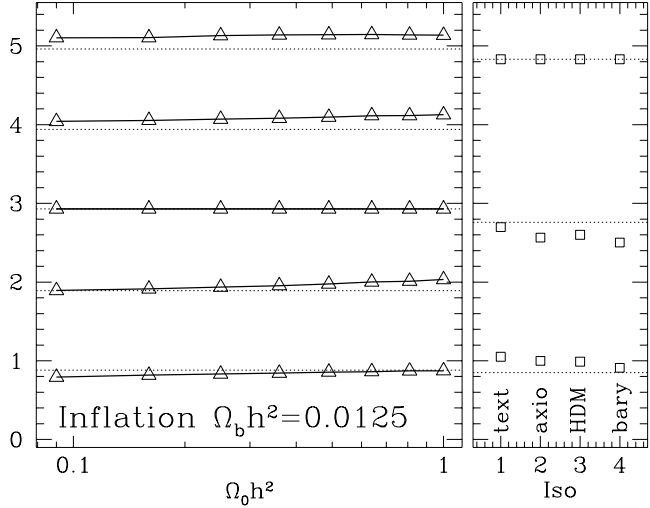}
\end{center}
\caption{Relative peak locations and the
harmonic series.}    
\mycaption{In the ideal case, inflationary
acoustic oscillations follow the cosine series
and driven isocurvature models a sine series (light dotted lines).
The ISW effect, baryon drag and diffusion damping serve to
distort the peak locations.
The isocurvature cases considered are:
baryon isocurvature (HSb);
textures (Crittenden \& Turok~1995);
axionic isocurvature (Kawasaki, Sugiyama \& Yanagida~1995) and
hot dark matter isocurvature (de Laix \& Scherrer~1995).
Numerical calculations (points) are normalized to the ideal
predictions of Eqs.~(37) and (38) at the third peak and are specifically
for $\Omega_0+\Omega_\Lambda=1$ though this constraint is irrelevant
for the peak rations. They demonstrate
that the two cases remain quite distinct especially in the ratio
between the first and third peak.
}
\label{fig:peak}
\end{figure}

\subsection{Robustness of the Harmonic Series}
\label{ss-robustness}

We have seen in \S\ref{ss-locationsd} that the harmonic series of acoustic peaks in a model
with super-horizon curvature fluctuations is given by the cosine series
$1:2:3\cdots$ for the locations of the peaks.
Compensated super-horizon fluctuations follow the sine series $1:3:5\cdots$.
Thus by measuring the ratio of the first three peak positions in 
$\ell$ space, these two possibilities may be distinguished.
There are two concerns that need to be addressed for this potential test of
inflation.  How robust is the harmonic prediction in the general class of
inflationary models? Can an isocurvature scenario where fluctuations are
generated inside the horizon mimic an inflationary series?  

Let us consider the first question.
Residual driving effects can distort the shapes and locations of the
first few peaks.  
Even for the pure adiabatic mode of \S \ref{ss-backreaction}, 
the peaks have not completely settled
into their asymptotic forms until $kr_s \gg 1$ and the peak
positions follow the series 
in $\Theta_0+\Psi$,
\begin{equation}
0.88 : 1.89 : 2.93 : 3.94 : 4.96 \cdots.
\label{eqn:Cosine}
\end{equation}
We will hereafter refer to these ratios as the ``cosine series''
despite the fact that it has not yet converged upon an actual cosine
oscillation by the first few peaks.  
In particular, the first peak is somewhat low in $\ell$ compared with the
expectation from the higher peaks.  Let us compare this prediction
against actual inflationary models from a numerical calculation
(see Fig.~\ref{fig:peak}, left panel, solid lines).  
All models do indeed exhibit the harmonic
structure predicted by Eq.~(\ref{eqn:Cosine}).  
There is a slight shift of the first few peaks downward
in $\ell$ as $\Omega_0 h^2$ is lowered.  As noted by HSb, if
the radiation is still dynamically important after last scattering,
the resultant ISW effect will shift power toward larger angles
and distort the first peak.
The even peaks are also shifted upwards as $\Omega_0 h^2$ increases
to make potentials deeper and baryon drag more important. 
Still, for reasonable matter content and a near BBN baryon content 
a ratio of 3rd to 1st peak of $\ell_3/\ell_1 \approx 3.3-3.7$
is a robust prediction
of inflation.  Furthermore the ratio of the first peak location
to the peak spacing is between $\ell_1/\ell_A \approx 0.7-0.9$.
The peak spacing is best measured by the prominent peaks, e.g.
for inflation $\ell_A = (\ell_3 - \ell_1)/2$. 
The only caveat is that models with extremely high
$\Omega_b h^2 \simgt 0.04$ {\it and} $\Omega_0 \gg \Omega_b $ 
may be dominated by the baryon drag effect.
In this case, only the compression phases are visible as peaks in the rms.
Thus every other oscillation is unobservable and the series becomes
$1:3:5\cdots$ like the isocurvature spectrum.  
For a model with reasonable matter content $\Omega_0h^2\simlt 0.25$, {\it no}
value of $\Omega_b h^2$ entirely eliminates the second peak.
For $\Omega_0 h^2=0.64$, the second peak disappears for
$0.04\simlt\Omega_bh^2\simlt0.5$ and the fourth for
$0.128\simlt\Omega_bh^2\simlt0.32$.
Even in this extreme case, only a gross violation of BBN would entirely mask
the inflationary pattern.
Thus the harmonic series is a robust prediction of inflation.

\subsection{Uniqueness of the Harmonic Series}
\label{ss-uniqueness}

Is the cosine harmonic series a unique prediction of inflation?
Just as in the inflationary case, for isocurvature models residual
driving effects create a downward shift from the
$ 1 : 3 : 5 \cdots$ pattern.  Even a pure driven isocurvature model with a
source $\Phi_s \propto x^{-1}$, which 
converges to the sine series for $k r_s \gg 1$ 
(see \S \ref{ss-driven}), gives
\begin{equation}
0.85 : 2.76 : 4.83 : 6.89 : 8.91 \cdots.
\label{eqn:Sine}
\end{equation}
We will hereafter refer to these ratios as the 
``sine series''.  How closely do real models follow this prediction?
For comparison, we show in the right panel of Fig.~\ref{fig:peak}
the ratios of the peak locations
for common isocurvature models found in the literature:
the baryon isocurvature (\cite{HSb}),
textures (\cite{CriTur}), 
axionic isocurvature (\cite{KawSugYan}), 
HDM isocurvature (\cite{deLSch}).
Notice that these models roughly correspond to the sine 
series prediction.   
Thus in all inflationary models with reasonable
baryon content, the ratio of peak locations 
should be distinguishable from the current models based on the isocurvature
restriction.  This suggests that to mimic inflation an isocurvature
model would have to either possess extreme conditions or
fine tuning.

\begin{figure}[t]
\begin{center}
\leavevmode
\epsfxsize=3.5in \epsfbox{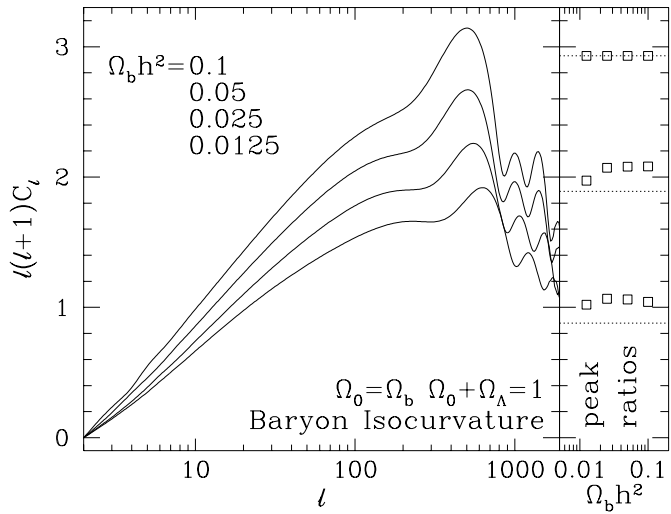}
\end{center}
\caption{Obscured isocurvature peak.}  
\mycaption{The first isocurvature ``peak''
appears as a shoulder and may be obscured especially in high
baryon models where the second peak is significantly higher (left
panel, arbitrary normalization).
If the second peak starts the harmonic series, the ratios (right
panel, points)
can be quite close to the cosine
prediction (dotted lines).  The points are normalized to the
cosine prediction at the third peak.  These
models can be distinguished by the peak to spacing ratio
and the morphology of the first compressional peak (2nd feature).
}
\label{fig:missing}
\end{figure}
 
Let us attempt to construct models which mimic inflation.  Since
the first inflationary peak appears
immediately after sound horizon crossing $k = \pi/r_s$,  
it is difficult to avoid or counter the driving effects 
which produce the sine series.  Can a series based on peaks arising
{\it after} sound horizon crossing somehow mimic the cosine series?
Consider the possibility that the first isocurvature peak in the
sine series Eq.~(\ref{eqn:Sine}) is missing or cannot be observed.
This is indeed likely in some models since the first peak can 
be quite low in intrinsic amplitude (see 
Fig.~\ref{fig:missing}, \S \ref{ss-heights} \& \S \ref{ss-first}).  
Furthermore, external effects such as tensor and vector modes
which naturally possess a feature on the horizon
scale at last scattering might mask or distort the peak.  
In this case, the observable peaks would follow the 
series: $2.8 : 4.8 : 6.9 \cdots$ which is close enough to the cosine
series to cause some concern, especially if only the first two peaks
are measured.  
More generally, one might have a source that turns on only after
sound horizon crossing so that the harmonic series is again shifted
toward smaller scales.  These possibilities however create
spectra which are not as close to the inflationary prediction as they 
initially might appear.

The crucial point is that although the starting point of the
harmonic series can change with the model, its spacing cannot.  
The separation between the peaks is fixed by the sound horizon
at last scattering.  Thus the harmonics cannot simply be scaled
to match the inflationary prediction.  The distinction is clearer
when we consider the ratio of the peak location to peak 
separation. The idealized inflationary prediction
of Eq.~(\ref{eqn:Cosine}) requires a ratio of the first peak
to the separation between peaks of $\ell_1/\ell_A \approx 0.88$.
If the second peak of the isocurvature prediction 
Eqn.~(\ref{eqn:Sine}) is to be taken for a first peak, it gives
a corresponding ratio of $\ell_1/\ell_A \approx 1.33$.  
In other words, we expect a 
factor of 1.5 difference between the two cases.  
The uncertainty displayed in the real models of 
Fig.~\ref{fig:peak} do not destroy the test. 
When comparing models with the same values for $\Omega_0 h^2$ the
idealized expectation is validated, e.g.~between
inflation and axionic isocurvature with $\Omega_0 h^2 =0.25$
and $\Omega_b h^2 =0.0125$ there is a 1.4-1.5 difference
in $\ell_1/\ell_A$ depending on which peaks
are used to measure the spacing.  
Even if the background parameters are unknown, to get the 
ratio as low as it is in the highest inflationary case ($\ell_1/\ell_A
\approx 0.9$ for
$\Omega_0 h^2 \simlt 1$), the first peak must be created 
around sound horizon crossing.  

The robustness of this test arises because the natural frequency
of the oscillator is related to the sound horizon.  Thus the
oscillations, if regular enough to be mistaken for inflation, must
appear regularly spaced with respect to that scale.
It is of course in principle 
possible that the external source is periodic and
drives the oscillator
at a different frequency.   It is amusing to note that even this
rather unlikely scenario can be distinguished.  The damping
tail provides a scale that is entirely independent on the nature
of the fluctuations.  If the background parameters are known,
the ratio of the damping scale to peak position provides an additional 
consistency test for the inflationary scenario.  
In models where the fluctuation is
generated inside the sound horizon, the ratio $\ell_1/\ell_D$ 
increases.  In
particular, for the above case of a missing first isocurvature
peak, the ratio increases by a factor of $1.5$ exactly as with
the case of the peak-to-spacing ratio.

\begin{figure}[t]
\begin{center}
\leavevmode
\epsfxsize=3.5in \epsfbox{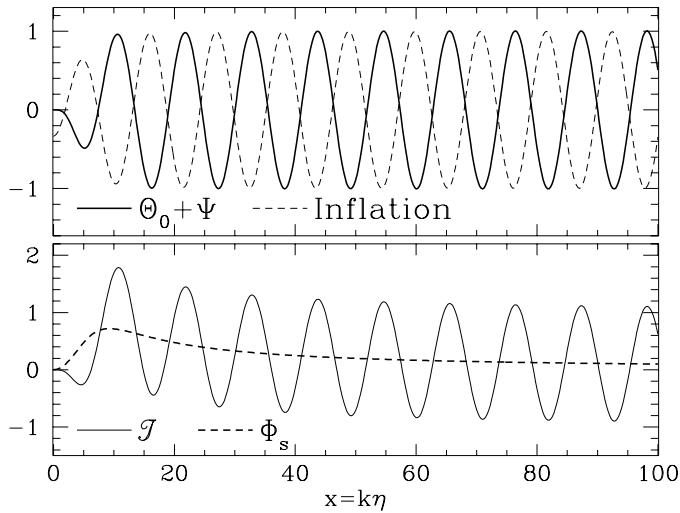}
\end{center}
\caption{Pathological isocurvature model.}
\mycaption{Here forced isocurvature fluctuations 
are tuned to match the locations of the
inflationary prediction (upper panel dotted lines) with
vanishing baryon content $R \rightarrow 0$.
Notice that even in this case the isocurvature oscillations
are out of phase with the inflationary prediction by 180 degrees.
With the inclusion of baryon drag, this leaves an observable signal in the
rms.}
\label{fig:pathol}
\end{figure}

Therefore, the only way to mimic the inflationary series is to tune the
behavior of the source at sound horizon crossing so that it immediately
generates a peak in the proper position but leaves no residual effects that
would distort the series of higher peaks.
Although contrived, this is possible. 
In Fig.~\ref{fig:pathol}, we plot a source given by Eq.~(\ref{eqn:SourceForm})
with $p_1 = 2$, $p_2 = 1$, and $\alpha = 0.05$ with $\Omega_b \rightarrow 0$.
In this case, we have constructed a source which dies off after
$x \approx \alpha^{-1}$ such as to leave the oscillations in a nearly pure
cosine mode.
The first five peaks follow a series $1 : 2.05 : 3.16 : 4.23 : 5.29$, very
close to the canonical inflationary prediction.  
Notice however that the prediction is 180 degrees out of phase with the
inflationary prediction.  This is an important fact that we will make use
of in the next two sections.
As we shall see, the crucial point is the effect of compensation near sound
horizon crossing.
One should also bear in mind that we have tuned the source to produce as
pathological a case as possible.   The cosine harmonic of peaks is essentially
but not entirely unique to the inflationary paradigm.
Its confirmation would strongly support the inflationary scenario. 

\subsection{First Peak}
\label{ss-first}

Can we employ additional information to eliminate the possibility
that an isocurvature model might mimic the inflationary model through 
the peak positions?  Compensation
from photon feedback discussed in \S \ref{ss-backreaction} and 
\S \ref{ss-heights} provides an essential 
distinction.  Near or above the horizon, the photons act to 
resist any change in curvature produced by the source if the universe
is radiation dominated (see Appendix for relaxations of this assumption).
Thus, the first peak in an isocurvature model, if it is sufficiently
close to the horizon to be confused with the inflationary prediction,
must be anti-correlated with the source.  In other words, the
first peak in the rms temperature represents the rarefaction
stage inside the potential well of the source rather than a compression
phase as in the inflationary prediction.  

Since the dilation effect $\dot \Theta_0 = -\dot\Phi$, which causes
$\Theta_0+\Psi \approx -2\Phi$, creates both the first peak and
the Sachs-Wolfe effect, the first 
feature in an isocurvature
spectrum is not truly a peak but a smooth turnover from the large scale
behavior as discussed in \S\ref{ss-driven} and \S \ref{ss-heights}.  
This point represents the epoch at which the fluctuation ceases to follow
the growth of the curvature and turns around to start the acoustic oscillation.
It creates the shoulder appearance of the first isocurvature feature
(see e.g.~Fig.~\ref{fig:damp} and Fig.~\ref{fig:missing}).
On the other hand, the inflationary fluctuation in the effective temperature
is proportional to $\cos(kr_s)$ and passes through a zero at $kr_s = \pi/2$
before the first peak (see Fig.~\ref{fig:drive}).  
The resultant spectrum thus exhibits a rather sharp break between the
gravitationally dominated Sachs-Wolfe tail and the first peak
(see Fig.~\ref{fig:damp}).  
The gravitational nature of the first peak in an isocurvature model makes it
substantially less prominent when compared with the low-$\ell$ tail.  The first
isocurvature peak also tends to be low with respect to the higher peaks because
it appears too close to horizon crossing to experience the full forcing effect
by the source.

Unfortunately these morphological distinctions are hard to quantify.
Indeed, the sharp rise to the first peak in the inflationary model
can be masked by the presence of the integrated Sachs-Wolfe effect
in a low $\Omega_0 h^2$ universe (\cite{HSa}).
In an isocurvature model, the prominence of the first feature can
be enhanced by a smooth bend in the power spectrum or vector
and tensor modes.  Nevertheless, the physical
distinction between the first adiabatic and isocurvature peaks
does suggest a robust way of distinguishing the models as
we shall now discuss. 

\begin{figure}[t]
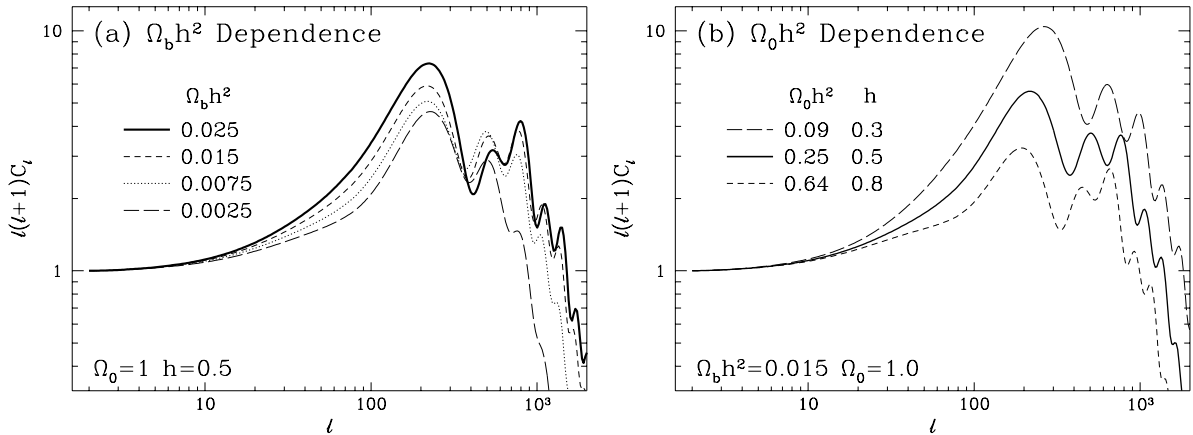

\begin{center}
\leavevmode
\epsfxsize=3.0in \epsfbox{f10a.epsf} \hskip0.4truecm
\epsfxsize=3.0in \epsfbox{f10b.epsf}
\end{center}
\caption{Baryon drag in the inflationary
model.}  
\mycaption{Baryon drag enhances the compressional, 
here odd, acoustic
peaks.  (a) Although diffusion damping at small scales coupled with the
intrinsically small baryon drag effect in low $\Omega_b h^2$ models
hides the effect, the third peak is clearly anomalously high in
all but the most extreme case $\Omega_b h^2 =0.0025$ which is
in clear violation of BBN constraints.
(b) Lowering $\Omega_0 h^2$ also reduces the effect by reducing
the potential fluctuation $\Psi$.}
\label{fig:dragheight}
\end{figure}

\subsection{Relative Peak Heights}
\label{ss-relative}

Whereas the inflationary
spectrum obeys a compression-rarefaction-compression pattern
an isocurvature model displays a rarefaction-compression-rarefaction
pattern.  
In the absence of baryons, there is no observable distinction 
between compression and rarefaction since only the rms
can be measured.  However, as we have seen in \S \ref{ss-drag}, baryons
enhance the compression at the expense of rarefaction leading
to an alternating series of peaks in the rms.  In fact, the example
in Fig.~\ref{fig:drag} is the same pathological model of 
\S \ref{ss-uniqueness} which has 
peaks at the inflationary locations by construction.  Notice
that for reasonable baryon content, the even peaks are enhanced
by the baryons, whereas the odd peaks are enhanced under
the inflationary paradigm.   Since this non-monotonic peak pattern
cannot be reproduced without introducing the appropriate features
in the initial $k$-spectrum of fluctuations, the 
pattern of anomalously high odd cosine peaks is a unique
feature of standard inflationary models.

Of course, an isocurvature model can also begin with a compression
if the first peak is so low in amplitude as to be unobservable
(see Fig.~\ref{fig:missing}).  However, such a model 
cannot simultaneously create the peak positions of the inflationary
prediction. In fact, regardless
of the peak positions, isocurvature models are still
unlikely to mimic a strongly alternating heights pattern. 
Since the first compressional isocurvature peak occurs well after horizon
crossing, neither the potentials of the source nor the photon-baryon
backreaction are large enough to cause significant baryon 
drag unless the baryon fraction is exceedingly high.  In this
case, the behavior of the potential implies an anomalously
high first compressional peak compared with the second 
due to baryon drag from the
residual self-gravity (see Fig.~\ref{fig:missing}).  

Although this pattern of peak locations and relative heights
is unique, it is not an entirely robust prediction of inflation.  
The intrinsic fractional effect at the $m$th peak is on the order
\begin{equation}
R(\eta_*) {\Psi(\eta_*,m k_A) \over \Psi(\eta_i,m k_A)} \approx R(\eta_*)
T(m k_A) \qquad (\eta_* \simgt \eta_{\rm eq})
\end{equation}
where $T(k)$ is the matter transfer function (\cite{BBKS}).
However, if the intrinsic effect is small, complications such as possible
smooth tilts in the initial spectrum and diffusion damping will make the
signal difficult to observe (see Fig.~\ref{fig:dragheight}).
For $\Omega_0h^2\approx0.25$ this occurs if $\Omega_bh^2\simlt0.007$
and for BBN baryon content for $\Omega_0h^2\simlt0.1$.  
Still, with high precision experiments, a sufficiently smooth initial
spectrum, and known thermal history, it is possible that even these extreme
cases may be testable (Hu \& White, in preparation).  On the other hand if
$R(\eta_*)T(m k_A) \gg 1$, baryon drag dominates
and the even $m$ peaks (rarefactions) no longer appear as peaks in the rms
(see Fig.~\ref{fig:drag}).  
In this case, the oscillations are small compared
with the offset and the distinction between compression and 
rarefaction stage is that they are maxima and minima of the rms  
respectively.  Since $T(k) \le T(0)$, 
a necessary but not sufficient 
condition for this to occur is $R > 1$ or a baryon content more than
3 times the BBN value.

\subsection{Summary}
\label{ss-threesumary}

In summary, inflationary models carry  an acoustic signature with 
distinct properties that  can
distinguish them from isocurvature models. 
The ratio of peak locations is a robust prediction of inflation.  
In particular, the ratio of third to first peak location should be
in the range $\ell_3 / \ell_1 \approx 3.3-3.7$ 
and the ratio of first peak location to
peak spacing should be between $\ell_1/\ell_A \approx 0.7-0.9$.
 If this pattern is {\it not} observed in the CMB, 
either inflation does not provide the main source
of perturbations in the early universe or BBN grossly
underestimates the baryon fraction.  The latter possibility is
treated more fully in \S \ref{ss-reconstruction}.  Observational
confirmation of the pattern would provide a plausibility proof for
inflation.  
It would thus require fine tuning for an isocurvature model  
to reproduce this spectrum.  To close this loophole in the test of
inflation, the relative peak heights can be observed.  Assuming the
location of the peaks follows the inflationary prediction, 
we find that
the high odd peak pattern of inflationary peaks is a unique
prediction of inflation with a near BBN baryon
content. Thus the locations and relative peak heights
can be used to ``prove'' the inflationary 
paradigm.  On the other hand, this signature relies on the baryon 
drag effect which may be small and difficult to observe in some
exotic inflationary models.  Moreover inflation does not preclude
the presence of isocurvature perturbations (\cite{Lin,YokSut}).
Disproof of inflation along these lines
is more difficult, a common problem in cosmology!

\begin{figure}[t]
\begin{center}
\leavevmode
\epsfxsize=3.5in \epsfbox{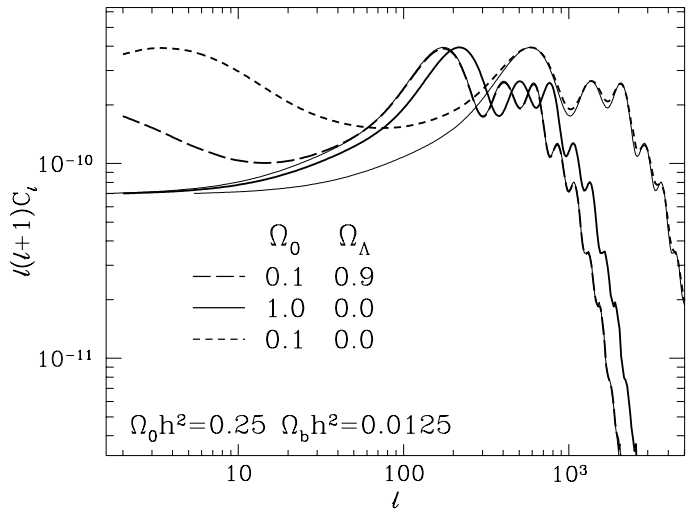}
\end{center}
\caption{Standard rulers and the angular diameter distance.}
\mycaption{Acoustic features
in the CMB, in particular the peak locations and the damping tail,
act as standard rulers with which a measurement of the curvature
can be made.  Thin solid
lines represent the $\Omega_0=1$, $\Lambda=0$ calculation scaled in $\ell$
to account for the projection in the $\Omega_0=0.1$ open and
$\Lambda$ models. }
\label{fig:angle}
\end{figure}

\section{Robustness of the Curvature Measurement}
\label{sec-curvature}

Acoustic oscillations in the CMB provide an interesting and unique
opportunity to measure the curvature of the universe.  Features 
in the spectrum supply standard rulers 
with which to make a classical test of 
the geometry (\cite{DZS,SugGou,KamSS,HuWhi}).  
If the inflationary scenario is correct, the curvature
can be measured to a few percent essentially from the location of the
first acoustic peak.
However, is the curvature measurement robust to changes in the underlying
model?  
Clearly, the location of the first peak does not contain enough
information.  Isocurvature models predict peak locations with
a different relation to the sound horizon.  Radical changes in
the baryon content from its BBN value and the thermal history from
standard recombination can shift their positions.
What additional information is required to ensure that the 
curvature measurement is robust?  

\subsection{Angular Diameter Distance}
\label{ss-angulardd}

It is instructive first to review the general 
case for measuring the curvature by the angular diameter distance
test.  For definiteness and simplicity, let us take the inflationary
example.  Here the acoustic spectrum in $k$
is purely a function of $\Omega_b h^2$ and $\Omega_0 h^2$.  The implied
fluctuations on the last scattering surface are viewed as anisotropies
on the sky today via the projection
\begin{equation}
\ell_{\rm feature}(K,\Lambda) = k_{\rm feature} r_\theta(K,\Lambda),
\label{eqn:Projection}
\end{equation}
where the comoving angular diameter distance to the last scattering
surface is 
\begin{equation} 
r_\theta(K,\Lambda)= |K|^{-1/2} \sinh[{|K|^{1/2}}(\eta_0 -\eta_*)].
\label{eqn:AngularDiameter}
\end{equation}  
for $K<0$ negatively curved universes.  For $K>0$, merely replace
$\sinh \rightarrow \sin$. 
The functional dependence arises since $\eta_0(K,\Lambda,\Omega_0 h^2)$ 
and $\eta_*(\Omega_0 h^2,\Omega_b h^2)$ and we are considering 
$\Omega_0 h^2$ and $\Omega_b h^2$ fixed for now. 
In general, negative curvature moves the acoustic features to smaller 
angles since a fixed physical scale then subtends a smaller
angle on the sky.  A cosmological constant moves features to larger angles
due to a decrease in the current horizon size from the rapid expansion.  

In Fig.~(\ref{fig:angle}), we display an example.  By holding $\Omega_bh^2$
and $\Omega_0 h^2$ constant, the acoustic features are fixed in physical
space and the shift in $\ell$ is entirely due to the projection.
Shifting the canonical $\Omega_0=1.0$ model by the angular diameter
distance scaling of an $\Omega_0=0.1$ model, we see that
Eq.~(\ref{eqn:AngularDiameter}) accurately accounts for the
effect.  
Notice that $K$ and $\Lambda$ are degenerate with respect to the acoustic
signature.  One cannot in principle measure them independently from these
features.  The degeneracy is broken at larger angular scales by the ISW effect
(\cite{HSb}).
Moreover for realistic $\Omega_0$, $\Lambda$ introduces little ambiguity in
the curvature measurement.  

It is clear that either the peaks or the damping tail can be
used as standard rulers to probe the curvature in this simple case.
We shall now consider the benefits and drawbacks of each in the more 
general
setting where the model for the perturbations and the background 
are not known {\it a priori}.
We then turn to what can be learned about the model for the
perturbations and how this additional information can be used to
refine the measurement of the curvature.

\subsection{Acoustic Peaks}
\label{ss-acousticpeaks}

The physical scale associated with the acoustic peaks is the sound horizon at
last scattering.  Unfortunately however, the exact relation between the 
peak locations 
and the sound horizon typically varies by a factor of 2 depending on
the model.  In this section, 
we discuss a how the peak spacing can be
used as a measurement of the spatial curvature which is far less
dependent on the model for the fluctuations.  

As discussed in \S \ref{ss-locationsd}, once the source has switched off inside the horizon the
acoustic peaks follow a phase shifted harmonic series.  
The key point is that regardless of the phase, the peaks are spaced
by $k_A=\pi/r_s$. Hence if the sound horizon at last scattering
is known, the spacing provides us with an angular diameter
distance test of the curvature. 
In making this measurement, the higher peaks form a better probe than the
lower peaks, since the latter can be shifted due to residual driving effects
and metric fluctuations between last scattering and the present. 
The peak spacings $\Delta\ell$ for the models of Fig.~\ref{fig:peak} are
shown in Table~\ref{tab:spacing} and should be compared with the prediction
$\Delta\ell=\ell_A \equiv k_A r_\theta$.  
As can be seen, the spacing becomes more regular and approaches the expected
value after the first peak separation $\ell_2 - \ell_1$. 
In the inflationary case shown here, baryon drag suppresses the even peaks and
distorts their locations.  For best accuracy, one should employ the spacings of
the prominent peaks for the test, e.g. $\ell_A=(\ell_3-\ell_1)/2$ in the
inflationary case.  
In general, three peaks will be necessary to assure accuracy of the test.

\begin{table}
\begin{center}
\begin{tabular}{l|cccc}
Model   & 1-2 & 2-3 & 3-4 & 4-5 \\ \hline
Inf     & 288 & 259 & 297 & 277 \\
HDM     & 204 & 282 & 272 & 295 \\
Tex     & 221 & 286 & 287 & 285 \\
Axi     & 204 & 295 & 275 & 298
\end{tabular}
\end{center}
\caption{The spacing between peaks for the suite of models in Fig.~7
with the same
background parameters $\Omega_0=1$, $\Omega_b=0.05$, $h=0.5$ and standard
recombination.  The spacing becomes more regular for the
higher peaks and converges toward the expected value of 290.  The spacing
yields a measure of the curvature of
the universe that is nearly independent of the model for the
fluctuations.}
\label{tab:spacing}
\end{table}

How well can this test measure the curvature given realistic uncertainties
in $\Omega_b h^2$ and $\Omega_0 h^2$ through $h$ which affect the
physical scale of the sound horizon at last scattering (see 
Fig.~\ref{fig:dampom})?  If the baryon
content is near or less than the BBN limit $\Omega_b h^2 \simlt 0.05$,
it has only a small effect since the photons dominate the fluid 
at last scattering.  The Hubble constant has a relatively
larger effect but even so a measurement of a peak spacing
$\ell_A \sim 290$ would require $\Omega_0 + \Omega_\Lambda
\simgt 0.7$ if $0.4 \le h \le 0.8$.  Note that the dependence on the
Hubble constant becomes weaker as it increases to make the
universe more matter dominated at last scattering.

The main drawback of this method is that it may be difficult to apply
in models where the forced effects continue well after horizon 
crossing.  In this case, the peak spacing may not become regular
until the higher peaks.  Near the diffusion scale, the power in the CMB 
fluctuations is exponentially damped.  
Foregrounds become more difficult to subtract on
these smaller scales and gravitational lensing by large scale structure 
(see e.g.~\cite{Sel95}) can wash out the oscillations to 
some extent.  It is even possible that
stochastic metric perturbations would leave the spectrum with no
distinct peaks
(see \S \ref{ss-forced}, \cite{Magetal}).  In these cases, we must turn
to the damping tail to measure the curvature. 

\subsection{Damping Tail}
\label{ss-dampingtail}

\begin{figure}[t]
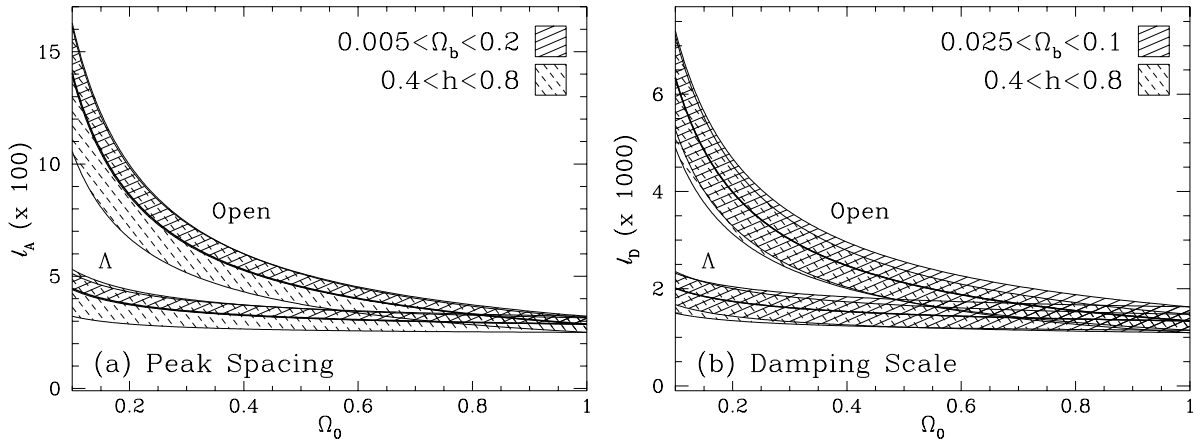

\begin{center}
\leavevmode
\epsfxsize=3.0in \epsfbox{f12a.epsf}\hskip0.4truecm
\epsfxsize=3.0in \epsfbox{f12b.epsf}
\end{center}
\caption{Peak spacing and damping scale as a function of $\Omega_0$.}
\mycaption{Even allowing for uncertainties in the baryon content (solid shading,
$h=0.5$) and Hubble
constant (dashed shading, $\Omega_bh^2 = 0.0125$),
open models with $\Omega_0 \simlt 0.5$ can be distinguished
from flat ($\Omega_0 + \Omega_\Lambda = 1$) $\Lambda$
models through either scale.  The damping scale is
entirely independent of the model for the fluctuations
but may be more difficult to measure
than the peak spacing. }
\label{fig:dampom}
\end{figure}

The diffusion damping length provides a 
standard ruler for the curvature measurement that is the most
robust against 
changes in the model for the fluctuations (see also \cite{HuWhi}).  
As discussed in \S \ref{ss-tail}, its location is dependent only on background parameters.
However, it is more sensitive to changes in $\Omega_bh^2$ and the thermal
history due to its dependence on the Compton 
mean free path at last scattering.
Fig.~\ref{fig:dampom}b displays the location of the damping tail $\ell_D$
defined as the multipole number at which the acoustic effect drops by
$e^{-2}$ in power.  It is related to the damping wavenumber
Eq.~(\ref{eqn:DiffusionLength}) by the simple projection of
Eq.~(\ref{eqn:Projection}). Here the thermal history has been 
fixed to follow standard recombination.  
By applying reasonable constraints on the other parameters, the damping scale
becomes a sensitive probe of the curvature.
Even allowing for a factor of 4 uncertainty in $\Omega_b h^2$,
an open model of $\Omega_0\simlt 0.5$ can be
distinguished from a flat $\Omega_0+\Omega_\Lambda=1$ universe.
An additional uncertainty arises from the Hubble constant 
which changes the age of the universe and expansion rate
at last scattering and hence the diffusion scale. 
In Fig.~\ref{fig:dampom}, we show the uncertainty from $h$ if $\Omega_b h^2$
is fixed by BBN.
Again open universes with $\Omega_0\simlt 0.5$ can be distinguished from the
flat cases for reasonable values of $h$ amounting to a factor of $4$ ambiguity
in $\Omega_0 h^2$. 

There is one caveat to this proposal for measuring the curvature.  It must
first be established that the damping tail is indeed due to 
photon diffusion
and not due to some intrinsic fall off in the source or secondary effect
between recombination and the present.
To test this assumption, one must measure the shape of the damping tail.
Diffusion damping leads to a near exponential fall off in the anisotropy
rather than the power law behavior expected in the alternate possibilities.
Thus it will be necessary to follow the damped oscillations into the diffusion
regime to establish its exponential character.  Even in a flat universe the
damping tail is at arcminute scales.  Detailed measurements of 
the primary signal may or may not be
observationally feasible due to possible foreground and secondary effects,
most notably the ISW effect.  Models with large small scale metric
fluctuations may also have their effective temperature at last scattering
dominated by baryon drag rather than acoustic oscillations.  This too
would mask the exponential signature.
Of course, if acoustic peaks in the angular power spectrum are also measured,
then it is almost assured that a break in the acoustic spectrum is due to
diffusion.  

\begin{figure}[t]
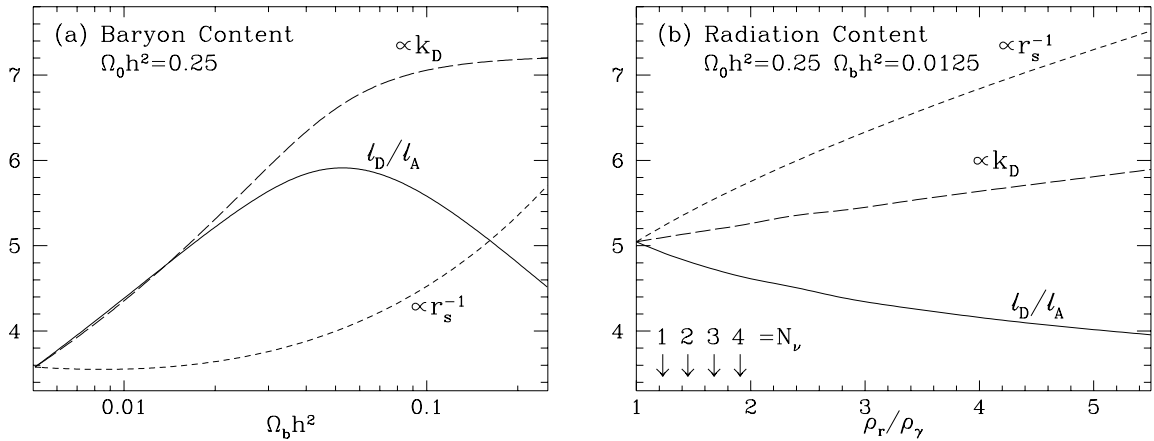

\begin{center}
\leavevmode
\epsfxsize=3.0in \epsfbox{f13a.epsf}\hskip0.4truecm
\epsfxsize=3.0in \epsfbox{f13b.epsf}
\end{center}
\caption{Exotic baryon and radiation content.}  
\mycaption{(a)
The sound horizon which forms the
basis of the peak-spacing test is independent of $\Omega_b h^2$ for
values near BBN or lower, whereas the damping tail is
insensitive to high $\Omega_b h^2$.  In either limit a robust
test exists.  The ratio of tail to peak-spacing $\ell_D/\ell_A$
can detect an exotic baryon content.  (b)
Changing the radiation content, e.g.~by altering the number of
relativistic neutrinos $N_\nu$, affects the expansion rate and thus
the two physical scales weakly.  Cases extreme enough to
affect the scales significantly can also be
distinguished by $\ell_D/\ell_A$.
}
\label{fig:ttp}
\end{figure}

\subsection{Exotic Baryon Content}
\label{ss-exoticb}

Thus far, we have mainly considered the robustness of the curvature measurement
to exotic sources of gravitational perturbations.
It is also possible that the background itself exhibits an exotic nature.
In particular, we have been implicitly assuming that the baryon content is
within a factor of several of its BBN value and the thermal history follows
standard recombination.  Let us now consider how the possibility of exotic
background properties may be handled, beginning with the baryon content.

If the baryon content is far from the BBN prediction of
$\Omega_bh^2\approx 0.01$--0.02, but {\it known} the approach to measuring the
curvature is unchanged since both the sound horizon and damping scale are
known as well.  It may be the case however that the prediction is violated by
observations that do not then constrain $\Omega_b h^2$ sufficiently.  If it is
low, the peak spacing still provides a good measurement of the curvature.
In this limit, the sound horizon at a fixed redshift becomes independent of
the baryon content and the only variation with $\Omega_b h^2$ is a weak
dependence on the redshift of last scattering (see Fig.~\ref{fig:ttp}a).
However, the peaks must still be measured for this test to work.  With a low
baryon content, the Compton mean free path and hence the diffusion length at
last scattering increases making the higher peaks difficult to observe.
The ratio of the damping tail to the peak-spacing $\ell_D/\ell_A$ provides a
rough model independent estimate of the number of peaks that are potentially
observable and is displayed in Fig.~\ref{fig:ttp}a (c.f. Fig.~\ref{fig:drag}).
Two peaks and hence the peak spacing is likely to be observable as long as
$\Omega_b h^2 \simgt 0.001$, which is close to the stellar mass density
$\Omega_{\rm stars} \approx 0.004$ for a reasonable Hubble constant.
Thus the peak spacing should provide a test of the curvature even under the
most extreme conditions. On the other hand, a curvature measurement from the
damping tail alone could fall victim to this exotic possibility.
However, the severely truncated acoustic spectrum in this case should prevent
such a misinterpretation of the position of the damping tail. 

In the case of an extremely high baryon content, the 
situation is reversed.  The damping tail now provides the more
robust estimate.  Since the effects of a
delay in last scattering and a decrease in the Compton mean 
free path tend to cancel, the physical scale is nearly independent
of the baryon content for high values (see Fig.~\ref{fig:ttp}).  
On the other hand, if the baryon content is raised more 
than a factor of 5 over the BBN value, the sound
speed and hence the sound horizon at last scattering
decreases.   Thus the peak spacing alone is an unreliable 
test of the curvature if the baryon content is extremely high 
$(\Omega_b h^2 \simgt 0.06)$ but unknown.

Is there a model-independent measure of the baryon content
that can test for anomalous values?  It should be clear 
from the discussion
above that the ratio of the damping tail to peak-spacing is
highly sensitive to $\Omega_b h^2$.  It is also a function of 
$\Omega_0 h^2$ but since $\Omega_0$ must be consistent with the
inferred curvature only uncertainties in the Hubble constant enter 
if $\Omega_\Lambda = 0$.    The ratio is double valued so that
its measurement would allow both a high and low baryon content
solution (see Fig.~\ref{fig:ttp}a).  As we shall see in 
\S \ref{ss-reconstruction}, baryon 
drag provides a distinction between the two extremes from a measurement
of the relative heights of the peaks.  In summary, an exotic 
baryon content is detectable and does not 
present a problem for the curvature
measurement if both the peak spacing and damping tail can be
measured.

\begin{figure}[t]
\begin{center}
\leavevmode
\epsfxsize=3.5in \epsfbox{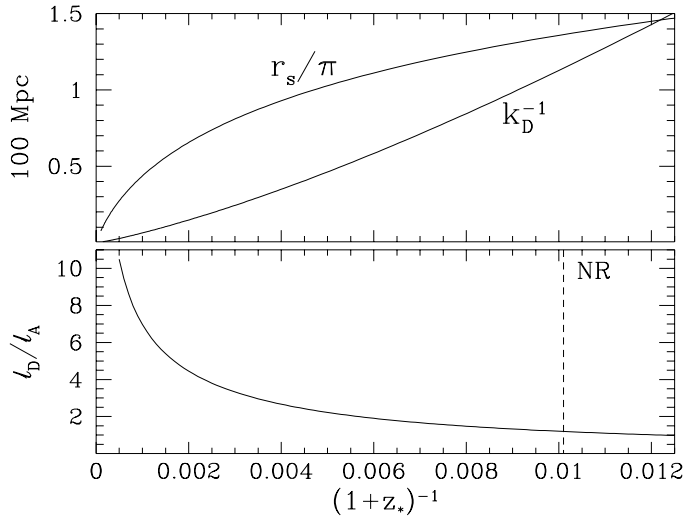}
\end{center}
\caption{Exotic ionization history}
\mycaption{We show the damping and peak-separation scales
in a model with instantaneous recombination at $z_*$.  For a gradual
recombination, a lower $\ell_D/\ell_A$ will always result.
The diffusion scale at last scattering by definition approaches the sound
horizon in the limit that no recombination (NR) occurred.
Assuming instantaneous recombination, the ratio $\ell_D/\ell_A$
roughly corresponds
to the number of observable peaks and can be used to discriminate against
exotic ionization histories.
Here $\Omega_0h^2=0.25,
\Omega_bh^2=0.0125$.  
}
\label{fig:therm}
\end{figure}

\subsection{Exotic Thermal History}
\label{ss-exotict}

Exotic thermal histories are another possibility.
By thermal history, we refer to both the 
thermodynamics of the expansion and the ionization history.
Massive decaying particles can create an epoch of matter domination before
last scattering which changes the age of the universe at last scattering and
hence both the sound horizon and 
the diffusion scale (\cite{WhiGelSil,BBE}).  
A less exotic example of the latter effect is provided by any model which
changes the epoch of equality. 

Let us take the a simple but illustrative example of a change in
the energy density in relativistic species.   This can arise for
example by changing the number of relativistic neutrinos or their
temperature.   In Fig.~\ref{fig:ttp}b, we plot the
sound horizon, damping scale and the ratio between the two
as a function of the fractional increase in the radiation density
$\rho_r/\rho_\gamma$.  The standard thermal history with three relativistic
neutrinos predicts $\rho_r/\rho_\gamma = 1.68$.  Raising the radiation
content increases the expansion rate and thus decreases both
the sound horizon and diffusion scale at last scattering.  Its
effects are relatively weak unless the standard prediction grossly
underestimates the radiation content.  Because the
diffusion scale is essentially the geometric mean of the horizon
and Compton mean free path, it is a weaker function of the radiation
content.  Hence, if the universe really has a sufficiently exotic
history to change the two physical scales at last scattering, it
should be detectable in the ratio of tail location to peak spacing 
$\ell_D/\ell_A$.

The use of $\ell_D/\ell_A$ as a probe of the thermal history is especially
powerful for models which delay last scattering significantly.  
This might occur due to 
energy injection near $z\sim10^3$ from
particle decays or non-linear fluctuations.
Since the
Compton mean free path grows as the universe expands, the diffusion scale can
approach the horizon scale (see Fig.~\ref{fig:therm}).  
Here full ionization is assumed
until a redshift $z_{*}$, when recombination is taken to occur instantaneously.
Instant recombination is not realistic; however it provides an upper limit
on $\ell_D$.
The ratio of $\ell_D/\ell_A$ in Fig.~\ref{fig:therm} then measures the delay
in recombination.  
Under the instantaneous recombination assumption it corresponds to the number
of observable peaks.
The actual observable value for any given model will be lowered in a manner
dependent on the details of recombination.

Even if recombination never occured the Compton mean free path will eventually
reach the horizon size due to dilution in the electron number density 
from the expansion.  At this point the
universe is effectively transparent and the photons free stream to the present.
Since by definition the diffusion length roughly corresponds to the horizon
scale at this epoch, $\ell_D/\ell_A\sim1$ and no acoustic oscillations are
apparent.
Thus recombination is a necessary condition for the acoustic signature to be
observable.  In fact for practical purposes, we require at least two observable
peaks.
For $\Omega_0h^2=0.25$ and $\Omega_bh^2=0.0125$, this translates to
recombination by $z_*=175$.  For comparison, late reionization after standard
recombination does not destroy the acoustic signal unless the optical depth
through the reionized epoch is greater than unity: $z_{\rm reion}\simgt100 
(\Omega_0 h^2/0.25)^{1/3}(\Omega_b h^2/0.0125)^{-2/3}$.

Although there is not enough information in the one number $\ell_D/\ell_A$ to
reconstruct all possible exotic thermal histories, it is possible to determine
that some exotic model is necessary if $\ell_D/\ell_A$ is not $\approx4-5$.
Further work on the damping tail is necessary before this constraint can be
tightened (Hu \& White, in preparation).
In this case, we trade precise knowledge of the curvature for evidence that
exotic physics is required in the early universe.
On the other hand, if the deviation is known to occur due to a specific cause
such as early energy injection or delays in equality, then the curvature can
once again be obtained.    
Moreover, with this information from both the damping tail and the peaks,
we cannot mistakenly infer a value of the curvature due to exotic
baryon content or thermal history. 

\subsection{Toward Reconstructing the Model}
\label{ss-reconstruction}

The previous sections 
considered constraints on the curvature that could be
made with minimal knowledge of the model for the perturbations.
However we
would ideally like to learn as much as possible about both questions from
the CMB spectrum.  In this section, 
we shall outline a program for measuring
the curvature by first establishing the basic model for structure formation.
Model properties such as the relative peak heights simplify the
task of measuring the curvature since they provide extra clues.
As noted in \S\S \ref{ss-acousticpeaks} -- \ref{ss-dampingtail}, this may be the only way to proceed if it proves
impossible to measure the higher acoustic peaks and damping scale.

We define the basic model possibilities by their dominant 
mechanism for forming anisotropies:

\begin{enumerate}
\item Inflation: correlated superhorizon curvature
fluctuations.
Although inflation may also produce iso\-curvature initial conditions, we class
them separately.	
\item Driven isocurvature (e.g.~axionic and baryonic isocurvature, 
textures, etc.): photon-compensated initial conditions. We allow for
the possibility that the first peak is obscured (see 
\S \ref{ss-uniqueness}) especially in high
baryon cases where it is strongly suppressed with respect to the second peak.
\item Forced isocurvature (e.g.~possibly strings): forces after but temporally correlated with horizon
crossing.  
\item Stochastic isocurvature (e.g.~possibly strings): 
random sub-horizon forces.
\item Reionized: any of the above in which the diffusion length and horizon
	length coincide, e.g.~fully ionized models.
\end{enumerate}
Each of these models can have anomalously
 high (low) baryon content defined as
$\Omega_b h^2$ significantly larger (smaller) than the big bang
nucleosynthesis (BBN) value of 1-2\%.
They can also be low in dark matter (CDM) such that $\Omega_0\sim\Omega_b$.
It is possible, though unlikely, that the model may be a mixture of
``inflation'' (as defined above) and driven isocurvature scenarios.  

For definiteness, we hereafter discount the possibility of an exotic
thermal history as treated in \S \ref{ss-exotict}, 
except for the case of late reionization
which is plausible in many models.
Furthermore, we assume that the Hubble constant and $\Lambda$ are sufficiently
constrained to make the confusion they introduce to the curvature measurement
irrelevant (see e.g.~Fig.~\ref{fig:dampom}).
We however relax the BBN constraints on $\Omega_bh^2$ as this can lead to
a qualitatively distinct acoustic pattern.

There are four tests that we can apply to fix the model and the 
curvature, based on the
acoustic signatures discussed in \S \ref{sec-signatures}.

\begin{enumerate}

\item Peak positions test: measure the locations and spacings 
	between the peaks in $\ell$.   Three useful items can be
	extracted from this test:

Peak ratios:  Ratios probe the nature of the model.
We define the distinguishing feature of a ``cosine'' series as a ratio of
third to first peaks of $\ell_3/\ell_1 
\approx 3-4$ and that of a ``sine'' of $\approx 5$.
Note that a true sine oscillation with missing first peak might be
observationally classed as a cosine.

Peak spacings: If they are regular, they provide an angular diameter
	distance test.  The physical scale corresponds to 
	$k_{A} = \pi/r_s$ if both compressional and
	rarefactional peaks are measured or $2\pi/r_s$ for the
	compressional ones.

Peak-to-spacing ratio:  Since the peak spacing is fixed by the sound  
	horizon, its ratio with respect to the first peak location
	provides a sensitive probe of the model.  Inflation predicts
	$\ell_1/\ell_A \approx 0.7-0.9$.

\item Relative height test: determine if any of the peaks are anomalously
high with respect to a smooth underlying spectrum.  Heights probe the baryon
content and forcing mechanism.

\item Tail-to-spacing test: measure the ratio of tail to peak-spacing
$\ell_D/\ell_A$.  If the peak spacing is regular,
the ratio measures the baryon content $\Omega_b h^2$ (and/or identifies
exotic thermal histories).  

\item Damping tail test: measure the shape and absolute location of the
damping tail as described in \S \ref{ss-dampingtail}.  
The shape confirms its acoustic nature
and the location provides an angular diameter distance test.

\end{enumerate}
As we shall see, in most cases not all four tests are necessary.  In contrast
to \S \ref{ss-dampingtail}, we shall here adopt the philosophy that information from the
damping tail only be used if all other tests are ambiguous.  
If the baryon content is assumed to be known
from nucleosynthesis, most of the tests involving the damping tail
are generally unnecessary.
On the other hand, the damping scale ideally should still be measured since
it provides a valuable consistency check on both the baryon content and the
 thermal history assumption.
Let us outline the program as a flow chart.  
It starts with the suite of peak position tests. There are 4 possibilities:

\noindent{{\it A. The peak ratios form a cosine series.}}

{Apply the relative heights test.}

\begin{enumerate}
\item High odd peaks forming a smooth sequence: 
	{\it inflation} with BBN or high $\Omega_b h^2$.  
	Tail-to-spacing fixes $\Omega_b h^2$. Spacing measures
	the curvature. 

\item High odd peaks with anomalously high first peak:
	high $\Omega_b h^2$ {\it driven isocurvature} with first peak obscured.
	Tail-to-spacing fixes $\Omega_b h^2$. Spacing measures
	curvature.  

\item High even peaks: tuned {\it forced isocurvature}
with near BBN $\Omega_b h^2$.
	Spacing measures the curvature

\item Monotonic: BBN or lower $\Omega_b h^2$.  Spacing measures
	curvature. 

\begin{itemize}

\item[a.] Peak-to-spacing ratio $< 1$: low CDM {\it inflation}.  
Tail-to-spacing can also measure $\Omega_b h^2$.

\item[b.] Peak-to-spacing ratio $> 1$: {\it driven isocurvature} with 
	obscured first peak. 
	Tail-to-spacing can also measure $\Omega_b h^2$.
	
\end{itemize}	
\end{enumerate}

\noindent{{\it B. The peak ratios form a sine series.}}

{Apply the relative heights test.}

\begin{enumerate}
\item High even peaks:
	{\it driven isocurvature} model with high $\Omega_b h^2$.  
	Tail-to-spacing fixes $\Omega_b h^2$. Spacing or
	damping tail measures curvature.

\item High second peak:
	{\it driven isocurvature} model with BBN $\Omega_b h^2$ or lower.
	Spacing measures curvature.
	Tail-to-spacing can also measure $\Omega_b h^2$.

\item Monotonic: high $\Omega_b h^2$, high CDM {\it inflation}.  Tail-to-spacing
	fixes $\Omega_b h^2$ and {\it half}-spacing determines curvature.

\end{enumerate}

\noindent{{\it C. Neither cosine nor sine}}

\begin{enumerate}
\item Peaks follow phase shifted harmonic:
	Coherent and tuned superposition of {\it inflationary} 
	and {\it driven isocurvature} (cf. \S \ref{ss-locationsd}).
	Tail-to-spacing fixes $\Omega_b h^2$.
	Spacing measures curvature. 

\item Peaks follow a cosine harmonic with gaps $1:3:4:5$:
	High $\Omega_b h^2$, relatively high $\Omega_0/\Omega_b$
	{\it inflation}.
	Tail-to-spacing fixes $\Omega_b h^2$.
	Spacing of higher peaks or {\it half}-spacing 
	across the gap measures curvature.
	Damping tail also measures curvature.

\item First few peaks irregular followed by a regular series.
	{\it Forced isocurvature} model with horizon crossing
	effects and BBN or lower $\Omega_b h^2$.  
	Peak spacing measures curvature. Tail-to-spacing ratio can 
	also measure the $\Omega_b h^2$.

\item Somewhat irregular peaks offset by smooth function 
	(e.g. Fig.~\ref{fig:drive}).  {\it Forced isocurvature} model 
	with effects well after horizon crossing and high 
	$\Omega_b h^2$. Take average value of {\it half}-peak spacing.  
	Tail-to-(average)-spacing fixes $\Omega_b h^2$.  
	Average spacing measures 
	curvature. 

\item Random locations:
	{\it Forced isocurvature} model with rapidly varying 
	metric well after horizon crossing.  Tail-to-spacing ratio
	constrains $\Omega_b h^2$.  Damping tail constrains curvature.  
\end{enumerate}

\noindent{{\it D. No peaks.}}

{Damping tail test.}

\begin{enumerate}
\item Exponential fall off:
	{\it Stochastic isocurvature} model. 
	Location of damping tail measures the curvature if $\Omega_b h^2$
	is known.
\item Power law fall off:
	{\it Reionized} or {\it stochastic isocurvature} model with peak
	source amplitude {\it far} inside horizon {\it and} 
	high $\Omega_b h^2$.  
	No robust constraints on the curvature are possible.
\end{enumerate}

\noindent
Thus, in all but the last case the acoustic signature constrains 
the curvature.  If the baryon content can also either
be measured from the signature itself or is known from external
constraints such as BBN, highly accurate measurements
of the curvature are possible.  Once the basic nature of the model and
background is determined through this program, detailed modeling
of the fundamental source for the fluctuations that formed large scale
structure may begin. 

\section{Conclusions}
\label{sec-conclusions}

We have generalized the formalism of \cite{HSa} to include
backreaction effects and examined the uniqueness and robustness of acoustic
signatures in the CMB.
By clarifying the role of compensation and feedback 
in the evolution of fluctuations, we
have shown that the phase of the oscillation, and hence the ratios of peak
locations, distinguishes inflation from ``typical'' isocurvature models.
Specifically, two robust test are the ratio of third to first peaks
and first to peak-spacing.
Our analysis also provides a better understanding of the structure of the
peaks in these models.
However since it is possible to imagine isocurvature models which 
are tuned to 
mimic the inflationary pattern of peaks, we
have stressed the importance of baryon drag, which allows us to distinguish
compressions from rarefactions in potential wells. 
This can help lift the confusion between
adiabatic models and these contrived isocurvature models.
Although the level of drag does not make this distinction clear for all
possible baryon densities, it is observable for the value
predicted by big bang nucleosynthesis.
Further understanding of the diffusion damping of anisotropies will allow us
to extend the lower limit of baryon densities for which we can distinguish
the rarefaction and compression peaks.

We focus on the importance of the damping tail as a measure of spatial 
curvature which is independent of the model for 
structure formation, and discuss the robustness of
curvature measurements from the CMB.  Even in the case where no acoustic
peaks are seen (e.g.~possibly string models) 
the damping scale can be estimated
under mild assumptions about the thermal history and baryon content.  
A more precise test is possible if the acoustic peaks are
regularly spaced as indeed expected of models without extreme  behavior
at small scales.
Either approach allows one to infer the physical scale of the
acoustic feature(s). Its projection onto the sky
allows us to perform a classical angular diameter distance test to determine
the curvature of the universe.    In the process of carrying out
these tests of the curvature, the general nature of the model for
the fluctuations can be reconstructed as well as the baryon content
of the universe.

All of these studies focus on the tale told by the CMB spectrum
taken as a whole.
In particular,  the acoustic pattern,
which arise from forced 
oscillations in the
photon-baryon fluid before recombination, leaves a distinct 
signature from which we may begin to reconstruct the cosmological 
mdoel.

\bigskip
\acknowledgments  
We would like to acknowledge useful conversations
with 
\href{http://www.sns.ias.edu/Main/faculty.html}{J. Bahcall}, 
P. Ferreira, A. Kosowsky, \& A. Stebbins.  We would also like
to thank R. Crittenden \& A. de Laix for supplying power spectra
from their calculations of the texture and HDM isocurvature models
as well as N. Sugiyama for use of his Boltzmann code. W.H. was supported
by grants from the NSF and W.M. Keck foundation.

\clearpage

\appendix

\section{Causality and Compensation}
\label{sec-causality}

In this Appendix, we clarify the role of causality in limiting the
behavior of fluctuations outside the horizon and its dependence on
gauge.  We show 
that only inflation can correlate the 
curvature fluctuations above the horizon 
assuming general relativity is the correct
description of gravity.
Moreover in the main text, we employed only the mechanism of
compensation, i.e.~the response of the photon-baryon fluid to 
a source, and {\it not} the full causal constraints which limit the
behavior of the source as well.
This proved sufficiently powerful to produce the distinctions in the acoustic
signature under the additional assumption that the compensation is provided by
the photons.
In this case, the feedback from the photon self-gravity produces the key
ingredient in making many of the signatures robust.  Since in the standard
FRW model, the universe is radiation dominated until near recombination, this
additional assumption is automatically satisfied.\footnote{Note that the
universe need only be radiation dominated when the fluctuation was well
outside the horizon, not at last scattering itself.  This is of course not
satisfied in models with significant reionization, but in this case acoustic
oscillations are not observable anyway.}
There are however exotic models where this assumption is
not satisfied.  For example, a decaying massive particle could cause the
universe to undergo a period of matter domination before recombination.
For this kind of situation, we need to examine the general case of compensation
and additional causal constraints on the model. 

Causality implies initial compensation in density fluctuations above the
horizon since the stress-energy tensor is conserved.  Heuristically, the
conservation law implies that changes in energy density at any location arise
from ``flows'' of energy density current across surfaces, or from displacements
of fluid elements.  Since fluid elements cannot be 
displaced ``beyond'' the horizon, this severely constrains the 
behavior of fluctuations at $k\eta\ll 1$.
However, because the stress-energy 
tensor obeys {\it covariant} conservation, the exact form of
causal constraints on
the density perturbation depends on the representation of the metric,
i.e.~the gauge.  

We have seen in Eq.~(\ref{eqn:Continuity}) that density perturbations
can also change due to the ``stretching'' effects from changes in 
the metric.  
A clever choice of gauge can eliminate such effects.
Let us examine the evolution of the density 
fluctuation in its
local rest frame.  
Note that this does {\it not} coincide with hypersurfaces with
zero bulk velocity for the total matter $V_T$
unless $g_{0i}$ vanishes.  
In the literature, this has been called the
comoving gauge (\cite{Bar}), velocity-orthogonal isotropic gauge
(\cite{KodSas84}) and total matter gauge (HSb). It 
represents the metric
fluctuations as
\begin{equation}
\begin{array}{rcl}
g_{00} \eal -a^2 (1 + 2 \xi Q),\\
g_{0j} \eal a^2 V_T k^{-1}Q_{|j},\\
g_{ij} \eal a^2 (1+2\zeta Q) \gamma_{ij},
\end{array}
\end{equation}
where $|$ represents a covariant derivative with respect to $\gamma_{ij}$
and recall $Q$ is the $k$th eigenfunction of the normal mode
decomposition.
These quantities are related to their Newtonian counterparts by
a gauge transformation $\tilde x^{\mu} = x^\mu + \delta x^\mu$.  
If the line element $ds^2$ is to remain
unchanged (to first order in $\delta x$),
\begin{equation}
\tilde g_{\mu\nu} = g_{\mu\nu} +
		g_{\alpha\nu}\delta x^\alpha_{\hphantom{\alpha},\mu} +
		g_{\alpha\mu}\delta x^\alpha_{\hphantom{\alpha},\nu} -
		g_{\mu\nu,\alpha}\delta x^\alpha.
\end{equation}
A similar relation follows for the stress energy tensor $T_{\mu\nu}$
and relates the matter quantities.  For densities, it is simpler to
note that it arises from a combination of a 
shift in time slicing and the 
background density evolution: 
$\tilde \delta_i Q = \delta_i Q - (\dot \rho_i / \rho_i) \delta x^{0}$.
These relations imply that from the Newtonian gauge, the rest frame is 
reached by the coordinate shift 
$\delta x^{\mu}=(V_T/k,\vec{0})Q$, and the perturbation quantities
are related by
\begin{equation}
\begin{array}{rcl}
 \xi \eal \Psi - \dot V_T/k 
 -\displaystyle{\dot a \over a}V_T/k, \\
\zeta \eal \Phi -\displaystyle{\dot a \over a}V_T/k,\\
\Delta_T \eal \delta_T + 3(1+w_T)\displaystyle{\dot{a}\over a}V_T/k ,
\label{eqn:NtoTMG}
\end{array}
\end{equation}
where $w_i = p_i/\rho_i$ and
subscript $T$ denotes quantities of the total
fluid from a sum over the particle constituents.
Here we have used the background evolution equation 
$\dot \rho_i / \rho_i
= -3 (1+ w_i) (\dot a / a) $.   
Note that the velocity $V_T$ is the same in the Newtonian and rest
frame
gauges since its transformation properties depend on $\delta x^i$. 
The familiar quantity $\zeta$ is the curvature of the spatial
hypersurfaces
in this gauge (\cite{Bar}).

Applying the gauge transformation Eq.~(\ref{eqn:NtoTMG}) to the 
Newtonian gauge equations or writing the Einstein equations
in this gauge
(see HSb, Eq.~(16), \cite{KodSas84} Eq.~4.7),
we obtain the evolution equation for the curvature,
\begin{equation}
\dot \zeta =  {\dot a \over a}\xi = - {\dot a \over a} {w_T \over 1+ w_T} 
	\left({\delta p_T \over p_T} - {2 \over 3}\Pi_T \right).
\label{eqn:ZetaEvol}
\end{equation}  
Unless $w_T = -1$ as in the case of the de Sitter phase, the
rest frame curvature remains constant in the absence of isotropic
(pressure) or anisotropic stress perturbations.  
The isotropic stress perturbation can be broken up into an
adiabatic and non-adiabatic part
\begin{equation}
{\delta p_T \over p_T } = c_T^2 w_T^{-1} \Delta_T + \Gamma_T ,
\end{equation}
with the sound speed of the total fluid given by
$c_T^2=\dot p_T/\dot \rho_T $.
The
adiabatic pressure perturbation is related to the curvature
fluctuation by a factor of $(k\eta)^2$ through the Poisson equation
and hence is a negligible source for $(k\eta)^2 \ll 1$.
The generalized compensation law is therefore that component 
evolution must balance at $k\eta \ll 1$ to keep the rest frame 
curvature constant in the absence of non-adiabatic
isotropic stress (``entropy'') or anisotropic stress
perturbations.  
Note that unlike the adiabatic isotropic 
stress, both of these sources are the same in any frame 
and admit no gauge ambiguity. 

Since the continuity equation in this gauge is
\begin{equation}
{d \over d\eta}\left( {\Delta_i  \over 1 + w_i} \right)
= -(kV_i + 3\dot\zeta),
\label{eqn:SimpleContinuity}
\end{equation}
for each of the individual particle species, 
the rest frame density perturbation obeys an ordinary conservation
law if these stresses are absent ($\dot \zeta = 0$).
In other words, the number density of each of the particle
constituents only changes through their bulk motion.  Thus we see that the
causal constraint is simplest in the rest frame.  Note that 
Eqs.~(\ref{eqn:SimpleContinuity}) and (\ref{eqn:ZetaEvol}) imply
that although density fluctuations
can change purely due to an evolving equation of state, this does not
affect the curvature fluctuation $\zeta$ above the horizon. 

Now let us consider the source of curvature fluctuations from
non-adiabatic pressure perturbations.  The analysis also applies
to anisotropic stress perturbations.
The curvature fluctuation generated by pressure perturbations 
is of order $\delta p_T/(p_T + \rho_T)$. 
This fact is somewhat counterintuitive since the physical mechanism
that converts a pressure fluctuation to a density fluctuation is
the movement of matter which is impossible beyond the horizon.
Let us examine its qualitative origin.  
Causality constraints in Fourier space do not require {\it no} evolution
for $k\eta \ll 1$.  In physical space, motion of matter through length scales
up to $\eta$ cause a suppressed evolution of the Fourier amplitude.  
A change in momentum density of the fluid is caused by the 
pressure gradient and generates a bulk velocity of order
$(k\eta)\delta p_T/(p_T + \rho_T)$. This then forms a kinematic density
fluctuation from the continuity equation 
(\ref{eqn:SimpleContinuity}) of order
$(k\eta)^2 \delta p_T/(p_T +\rho_T)$ or a curvature fluctuation of order
$\delta p_T/(p_T+\rho_T)$.  
Thus the residual curvature fluctuation induced by motion of matter inside
the horizon is generically of order
$w_T/(1+w_T)$ times the pressure perturbation.

Causality implies that before matter has had a chance to move around,
the universe must obey the isocurvature condition $\zeta = 0$
or $\Delta_T = 0$.\footnote{More precisely $\Delta_T=0$ aside from
a $\zeta = 0$ decaying mode [see Eq.~(\ref{eqn:DecayingMode})].}   
If the condition is met by balancing perturbations
in the different components of the fluid, a non-adiabatic pressure
fluctuation generically arises
\begin{equation}
p_T \Gamma_T \equiv \delta p_T - c_T^2 \delta\rho_T 
 =  \sum (c_i^2 - c_T^2) \delta \rho_i,
\end{equation}
if the equation of state of the balancing components differ.
As we have seen, $\Gamma_T$ {\it can} produce a curvature
fluctuation even at $k\eta\ll1$.  How does this possibility 
affect our arguments concerning the uniqueness of the inflationary 
spectrum?   
First, we need to relate
the rest frame curvature to the Newtonian curvature.
Employing the continuity equation (\cite{KodSas84}, Eq.~4.7)
in Eq.~(\ref{eqn:NtoTMG}), we find (\cite{Lyt,MukFelBra})
\begin{equation}
\zeta = \Phi + {2 \over 3}{1 \over 1 + w_T} \left( \Phi + {a \over \dot a}
\dot \Phi \right),
\end{equation}
in the absence of anisotropic stress.
If the equation of state is constant and $\Phi$ evolves as 
a power law, $\zeta \propto \Phi$.  The two curvature fluctuations
are comparable except in the degenerate 
case where $\zeta = 0$ and
\begin{equation}
{\dot \Phi \over \Phi} = -{\dot a \over a}[{3 \over 2}(1+w_T)+1].
\label{eqn:DecayingMode}
\end{equation}
In fact, this is the case of the source-free decaying mode
described by Eq.~(\ref{eqn:PureModes}).   The decaying mode thus
carries no curvature perturbation in the rest frame.  The other
case in which $\zeta$ and $\Phi$ behave differently is through
a change in the equation of state.  For example, through the
matter radiation transition $w_T$ goes from $1/3 \rightarrow 0$.
Although $\zeta$ remains constant in the absence of stress perturbations,
$ \Phi $ drops by a factor of $9/10$ through the transition.
For most purposes however, we can think of $\zeta$ and $\Phi$
as nearly interchangeable.
 
The mechanism by which non-adiabatic pressure perturbations generate
curvature fluctuations 
is of course already implicitly encorporated in our analysis and
is the cause of the curvature not being strictly zero outside at $k\eta\ll1$
but merely suppressed.
In the photon dominated limit, only a small pressure fluctuation is needed to
compensate a rather large density fluctuation in the source.  Thus, the
curvature generated by this effect is negligible.  In the matter dominated
limit, pressure fluctuations cannot move a large amount of energy density as
exhibited by the $w_T/(1+w_T)$ suppression.  However, as the universe changes
from radiation to matter domination a relatively 
significant curvature fluctuation can be
generated.

The baryon isocurvature model provides a concrete example.  In this case,
the non-adiabatic pressure perturbation is
\begin{equation}
\Gamma_T = -{1 - 3w_T \over 1+ w_T} S,
\end{equation}
where recall that $S=\delta(n_b/n_\gamma)=\delta_b-3\Theta_0$ is the entropy
fluctuation.
Notice that the pressure fluctuation is small as long as the universe is
radiation dominated $w_T=1/3$.  The evolution equations may be exactly solved
in the $k\eta\ll1$ limit such that the Newtonian curvature is
$\Phi={1\over8}(a/a_{eq})S$ in the radiation-dominated limit and ${1\over5}S$
in the matter dominated limit (see HSb, Eq.~27).
Notice that the curvature is a constant in the matter dominated limit even
outside the horizon.  Around equality, 
pressure perturbations of order $S$ generate curvature
fluctuations of the same order.
Since in the matter dominated limit, pressure is no longer effective, this
curvature fluctuation is frozen in.

If one considers the evolution of a single $k$-mode, distinguishing between
this and the inflationary case would be difficult since they both exhibit a
constant curvature fluctuation above the horizon.
In the standard scenario, this is not a problem since the matter-radiation
transition cannot occur early enough without overclosing the universe
$(\Omega_0h^2\gg1$).  Note that the curvature fluctuation must be constant
well before horizon crossing for all observable scales in order to mimic
inflation.  However it is possible with decaying massive particle scenarios
for the universe to undergo a period of matter-domination before the ordinary
radiation-dominated epoch.

In this case, the spectrum of non-adiabatic pressure 
perturbations implied by causality serves to
distinguish the model from inflation.  
Causality forbids spatial correlations in the spectrum of such 
perturbations, so
$|\Gamma_T(k)|^2$ and hence $|\Phi(k)|^2$ are constant in $k$, 
i.e.~white noise (or steeper if symmetries can be imposed),
before horizon crossing.
Notice that this agrees with the familiar result that the density
perturbations generated by the motion of matter have a $|\Delta_T|^2
\propto k^4$ tail for $k\eta \ll 1$ (\cite{Zel,RobWan}).
This translates to a steeply rising spectrum of acoustic fluctuations compared
with the inflationary case of an approximately scale invariant spectrum
$|\Phi(\eta_i,k)|^2\propto k^{-3}$.
Thus acoustic modes associated with a constant curvature perturbation
outside the horizon generated from an isocurvature initial condition
are both easily distinguished from inflation and observationally ruled
out!  The most general isocurvature spectrum with scale invariant curvature
perturbations at horizon crossing is $|\Phi(\eta,k)|^2=F(k\eta)k^{-3}$.
With the requirement of white noise perturbations outside the horizon,
$\Phi$ must grow as $\eta^{3/2}$ before horizon crossing.  Since only models
like these, with curvature fluctuations growing until horizon crossing, require
observation of the acoustic signature to distinguish them from inflation,
our assumption in the main text is justified.  It is of course still
possible that tuned effects around horizon crossing can mimic an 
inflationary spectrum in such a 
exotic scenario where the regulatory effects
of photon feedback are absent.  However, since both tuning and a drastic
modification of the thermal history is necessary, we do not consider
this possibility to be worth considering.

\vskip 1truecm
\noindent
{\tt whu@sns.ias.edu}

\noindent
{\tt http://www.sns.ias.edu/$\sim$whu}

\end{document}